\documentclass[%
	aps,prd,10pt,twocolumn,%
	tightenlines,notitlepage,%
	superscriptaddress,preprintnumbers,%
	nofootinbib,longbibliography%
]{revtex4-2}

\usepackage[utf8]{inputenc}
\usepackage[T1]{fontenc}
\usepackage{lmodern}

\usepackage{amssymb,amsmath,amsfonts}
\usepackage{xifthen,xspace,siunitx}
\usepackage{xcolor}
\usepackage{graphicx}
\usepackage{hyperref}
\usepackage[capitalise]{cleveref}
 
\newcommand*{\Lbr}{\ensuremath{\Lambda_\text{br}}\xspace}
\newcommand*{\mrel}{\ensuremath{m_\phi}\xspace}
\newcommand*{\frel}{\ensuremath{f_\phi}\xspace}
\newcommand*{\rX}{\ensuremath{r_{X}}\xspace}
\newcommand*{\vEW}{\ensuremath{v_H}\xspace}
\newcommand*{\Mpl}{\ensuremath{M_\text{Pl}}\xspace}

\newcommand*{\srel}{\ensuremath{\sin\theta_{h\phi}}\xspace}

\newcommand*{\tarh}{\ensuremath{\tau_\text{rh}}\xspace}
\newcommand*{\arh}{\ensuremath{a_\text{rh}}\xspace}

\newcommand*{\tapp}{\ensuremath{\tau_\text{pp}}\xspace}
\newcommand*{\app}{\ensuremath{a_\text{pp}}\xspace}
\newcommand*{\Hpp}{\ensuremath{H_\text{pp}}\xspace}
\newcommand*{\ra}{\text{ra}}
\newcommand*{\tara}{\ensuremath{\tau_\ra}\xspace}
\newcommand*{\ara}{\ensuremath{a_\ra}\xspace}
\newcommand*{\Hra}{\ensuremath{H_\ra}\xspace}
\newcommand*{\Tra}{\ensuremath{T_\ra}\xspace}
\newcommand*{\Tosc}{\ensuremath{T_\text{osc}}\xspace}

\newcommand*{\kpp}{\ensuremath{k_\text{pp}}\xspace}
\newcommand*{\kra}{\ensuremath{k_\text{ra}}\xspace}
\newcommand*{\km}{\ensuremath{k_\text{m}}\xspace}
\newcommand*{\ke}{\ensuremath{k_\text{ex}}\xspace}

\newcommand*{\tat}{\ensuremath{\tilde{\tau}}\xspace}
\newcommand*{\kt}{\ensuremath{\tilde{k}}\xspace}

\newcommand*{\freq}{\ensuremath{f}\xspace}

\newcommand*{\rhoGW}{\ensuremath{\rho_\text{GW}}\xspace}
\newcommand*{\rhoc}{\ensuremath{\rho_c^0}\xspace}
\newcommand*{\OGW}{\ensuremath{\Omega_\text{GW}}\xspace}
\newcommand*{\fp}{\ensuremath{\freq_\text{peak}}\xspace}
\newcommand*{\Ph}{\ensuremath{\mathcal{P}_{h'}}\xspace}
\newcommand*{\SNR}{\ensuremath{\varrho}\xspace}
\newcommand*{\Oeff}{\ensuremath{\Omega_\text{eff}}\xspace}
\newcommand*{\Neff}{\ensuremath{N_\text{eff}}\xspace}
\newcommand*{\Ics}[1]{\ensuremath{I_{#1}}\xspace}
\newcommand*{\Icst}[1]{\ensuremath{\tilde{I}_{#1}}\xspace}
\newcommand*{\normX}{\ensuremath{\mathcal{A}_X}\xspace}
\newcommand*{\normk}{\ensuremath{\mathcal{A}_k}\xspace}
\newcommand*{\order}[1]{\ensuremath{\mathcal{O}\left(#1\right)}\xspace}

\newcommand*{\vect}[1]{\ensuremath{\mathbf{#1}}\xspace}
\newcommand*{\hc}{\ensuremath{\text{h.c.}}\xspace}

\newcommand*{\experiment}[1]{{#1}}
\newcommand*{\muAres}{$\mu$\experiment{Ares}\xspace}

\DeclareMathOperator{\CosInt}{Ci}
\DeclareMathOperator{\SinInt}{Si}

\def\bea  {\begin{equation}\begin{aligned}}   \def\eea  {\end{aligned}\end{equation}} 
 
\DeclareSIUnit{\year}{yr} 
\binoppenalty=\maxdimen
\relpenalty=\maxdimen 

\Crefname{equation}{Eq.}{Eqs.}
\crefname{section}{Sec.}{Secs.}
\Crefname{section}{Section}{Sections}

\begin{document}

\preprint{} 

\title{Gravitational wave echo of relaxion trapping}

\author{Abhishek~Banerjee}
\email{abhishek.banerjee@weizmann.ac.il}
\affiliation{Department of Particle Physics and Astrophysics, Weizmann Institute of Science, Rehovot 7610001, Israel}
\author{Eric~Madge}
\email{eric.madge-pimentel@weizmann.ac.il}
\affiliation{Department of Particle Physics and Astrophysics, Weizmann Institute of Science, Rehovot 7610001, Israel}
\author{Gilad~Perez}
\email{gilad.perez@weizmann.ac.il}
\affiliation{Department of Particle Physics and Astrophysics, Weizmann Institute of Science, Rehovot 7610001, Israel}
\author{Wolfram~Ratzinger}
\email{w.ratzinger@uni-mainz.de}
\affiliation{PRISMA$^+$  Cluster of Excellence and Mainz Institute for Theoretical Physics, Johannes  Gutenberg-Universit\"at  Mainz, 55099 Mainz, Germany}
\author{Pedro~Schwaller}
\email{pedro.schwaller@uni-mainz.de}
\affiliation{PRISMA$^+$  Cluster of Excellence and Mainz Institute for Theoretical Physics, Johannes  Gutenberg-Universit\"at  Mainz, 55099 Mainz, Germany}

\date{\today}

\begin{abstract}
	To solve the hierarchy problem, the relaxion must remain trapped in the correct minimum, even if the electroweak symmetry is restored after reheating. 
	In this scenario, the relaxion starts rolling again until the backreaction potential, with its set of local minima, reappears. 
	Depending on the time of barrier reappearance, Hubble friction alone may be insufficient to retrap the relaxion in a large portion of the parameter space. 
	Thus, an additional source of friction is required, which might be provided by coupling to a dark photon.
	The dark photon experiences a tachyonic instability as the relaxion rolls, which slows down the relaxion by backreacting to its motion, and efficiently creates anisotropies in the dark photon energy-momentum tensor, sourcing gravitational waves. 
	We calculate the spectrum of the resulting gravitational wave background from this new mechanism, and evaluate its observability by current and future experiments. 
	We further investigate the possibility that the coherently oscillating relaxion constitutes dark matter and present the corresponding constraints from gravitational waves. 
\end{abstract}

\maketitle


\section{Introduction}
\label{sec:introduction}

\begin{figure*}
	\centering
	\includegraphics[width=2\columnwidth]{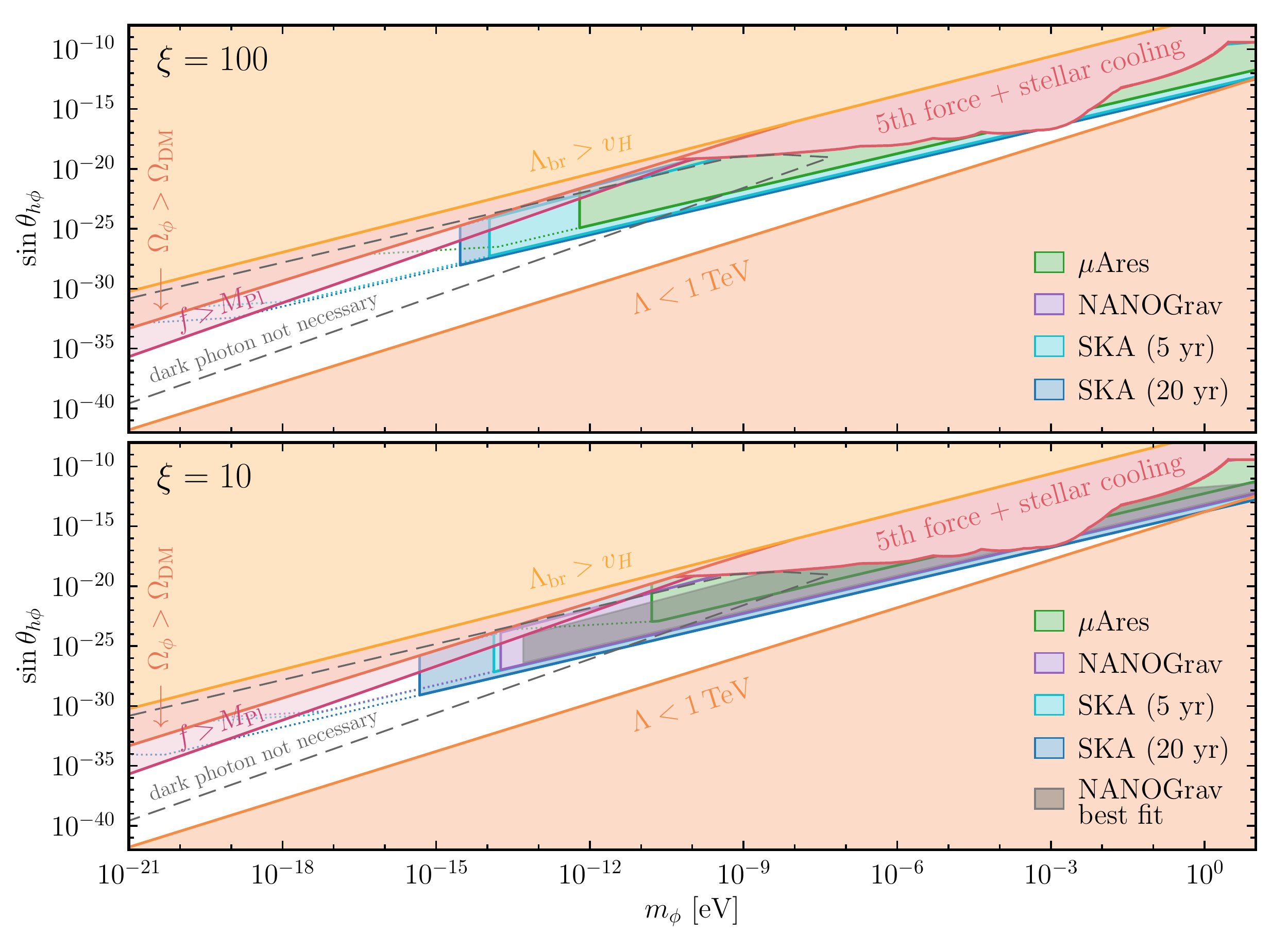}
	\caption{%
	Available parameter space for $\xi=100$ ({\bf top}), and for $\xi=10$ ({\bf bottom}). 
	The red and orange shaded regions are excluded by the indicated constraints or combinations thereof.
	Above the red solid line, the relaxion decay constant becomes super-Planckian. 
	The grey dashed line encloses the parameter space in which relaxation can be realized without dark photon friction, which is discussed in more details in \cref{app:minimalScenario}.
	The prospective GW sensitivity of \muAres~(green) as well as \experiment{SKA} after an observation period of 5~years~(turquoise) and 20~years~(blue) is indicated by the respective coloured regions.
	In the purple coloured region, a subrange of the viable reappearance temperatures can be excluded using current \experiment{NANOGrav} data from the 11~year data set. 
	The regions bounded by the coloured dotted lines need super-Planckian decay constants to be probed by the respective experiment.  
	In the lower panel, the grey shaded region can reproduce our best-fit spectrum (at $\Tra \sim \SI{20}{\MeV}$) to the potential stochastic GW background seen in the recent \experiment{NANOGrav} data.
	An animated version of the lower panel of the figure, explicitly showing the dependence on the reappearance temperature, can be found in the ancillary material of this paper.
	}
	\label{fig:RelaxionGW}
\end{figure*}

The weakness of the gravitational force ensures that gravitational waves (GWs) can store information about the history of our Universe. 
On the other hand, though, it also implies that only dramatic and cosmological events may leave imprints that can be observed at present or in the near future. 
GW~detectors are, for instance, sensitive to primordial phase transitions, inflation, and oscillatory motion of ultralight fields (see, e.g., Refs.~\cite{Caprini:2018mtu,Caprini:2019egz} for a recent discussion of potential signals of cosmological origin).

In this work, we introduce a different mechanism that leads to the production of GWs in the early Universe.
In the following, we demonstrate the idea using the relaxion framework~\cite{Graham:2015cka}; however, the core scheme may be applicable to other forms of new physics. 
At early times, the relaxion is trapped in a local minimum. 
However, assuming that the reheating temperature is above the electroweak (EW) scale,%
\footnote{%
	The requirement of a high reheating temperature, motivated by a large class of inflation models (see, e.g., Ref.~\cite{Salopek:1988qh} and references therein), is generic and also needed in most models that explain the observed baryon abundance (see ,e.g., Ref.~\cite{Bodeker:2020ghk} and references therein).
} 
we expect the EW symmetry to be restored.
Hence, the minimum in which the relaxion was originally trapped disappears, and the relaxion starts rolling down its potential. 
At some temperature below the EW scale, we expect the set of local minima to reappear.
There is a limited region of the model's parameter space such that the relaxion would be trapped still within the same minimum~\cite{Banerjee:2018xmn, Banerjee:2020kww}. 
In this case, even the minimal realization of the relaxation mechanism could lead to a viable ultra light dark matter (DM) candidate~\cite{Banerjee:2018xmn}, which may be tested in the future due to the presence of relaxion-Higgs mixing~\cite{Choi:2016luu, Flacke:2016szy}.

Despite the attractiveness of this minimal setup, there is a sizeable region of parameter space where the post-reheating displacement of the relaxion would be too large and the relaxion would not be trapped but instead would descend in an uncontrolled way toward the global minimum. 
This runaway can be avoided if an additional source of friction is added to the model. Coupling a scalar field to a dark $U(1)$ gauge field strength is known to lead to an efficient energy loss mechanism in the form of tachyonic dark photon field production~\cite{Choi:2016kke,Hook:2016mqo} (see Refs.~\cite{Fonseca:2019ypl,Fonseca:2019lmc} for the case of a relaxion or other axionlike particles (ALPs) in the mass range $\gtrsim\mathcal{O}(\si{\MeV})$, where self-friction may be induced, or Ref.~\cite{Wang:2018ddr} for stopping via an instability in a modified relaxion potential, albeit outside the region of interest of this work).
Dark photon production quickly reaches a quasi-steady-state where the friction balances the slope of the potential.
At each time, the dominantly produced momentum-mode $k = \xi a H$ is about to exit the tachyonic band, where $\xi$ is an $\mathcal{O}(10-100)$ parameter, $H$ is the Hubble rate, and $a$ is the scale factor.  

The energy density stored in this mode is roughly constant, $\rho_X\sim \mrel^2 \frel^2$~\cite{Banerjee:2018xmn}, where $\mrel$ and $\frel$ are the relaxion mass and the decay constant, respectively, and is the source of GW production. 
The rolling of the relaxion and the dark photon production stop around the time when the potential barriers reappear, which traps the relaxion and ends this epoch.%
\footnote{%
	For simplicity, we are agnostic about the details of the reappearance process of the backreaction potential and assume instantaneous reappearance. 
	Various considerations regarding barrier reappearance can also generate gravitational waves, a scenario considered in Ref.~\cite{Barducci:2020axp}.
}

Given the above mechanism, we can estimate the GW signal as follows.
First, the peak frequency at present time is obtained by redshifting the dominant $k$-mode at the time of reappearance, cf.\ \cref{eq:fPeak},
\begin{equation}
	\fp \sim \frac{\ara}{a_0} \xi \Hra \sim \SI{1}{\micro\Hz}  \left(\frac{\xi}{25}\right) \!\left(\frac{\Tra}{\SI{1}{\GeV}}\right),
	\label{eq:peakestimate}
\end{equation}
where $\Tra$ is the temperature when the potential barriers reappear. 
Here and in the following, quantities indexed ``0''   and ``\ra'' are evaluated today and at the time when the potential barriers reappear, respectively. 
Second, the GW amplitude is roughly given by the square of the energy density stored in the dark photon just before $\Tra$, cf.\ \cref{eq:rhoGWtot},
\begin{align}
\!\!\OGW^\text{peak} 
	& \sim \frac{1}{\rhoc}\, \frac{(c_\text{eff}\,\rho_X^\ra/\fp^\ra)^2}{\Mpl^2 } \left(\frac{\ara}{a_0}\right)^4
	\nonumber\\
	& \sim  \num{e-10} 
	\left(\frac{25}{\xi}\right)^{\!2} \!
	\left(
		\frac{\mrel}{\SI{0.1}{\eV}}
		\frac{\frel}{\SI{e10}{\GeV}}
	\right)^{\!4} 
	\!\left(
		\frac{{\SI{1}{\GeV}}}{\Tra}
	\right)^{\!8} \!
	\label{eq:AmplitudeEstimate}\\
	& \sim  \num{e-12}
	\left(\frac{25}{\xi}\right)^{\!2} \!
	\left(		
		\frac{\SI{0.1}{\eV}}{\mrel}
		\frac{\sin\theta_{\phi h}}{\num{e-13}}
	\right)^{\!\!12}\!
	\left(
		\frac{\si{\TeV}}{\Lambda}
		\frac{\si{\GeV}}{\Tra}
	\right)^{\!8}\nonumber\!\!,\!\!
\end{align}
where $\rhoc = 3 \Mpl^2 H_0^2$ is the critical energy density of the Universe today, $\fp^\ra = \xi \Hra$ is the peak frequency without red-shifting, and $c_\text{eff}^2 \simeq 1/4$ is the efficiency factor for converting dark photon energy into GWs obtained from the analytic calculation in \cref{app:GWcalculation}.
Note that $c_\text{eff} \sim \order{1}$ as the relaxion potential energy at each time is predominantly deposited in a narrow range of exponentially growing dark photon momentum modes, resulting in large time-dependent inhomogeneities and an efficient generation of anisotropic stress sourcing GWs, in a similar manner as in the case of tachyonic preheating~\cite{Khlebnikov:1997di,Easther:2006gt,Dufaux:2008dn,Adshead:2018doq}.
In the last line, we have reexpressed the signal strength in terms of the relaxion-Higgs mixing $\sin\theta_{h\phi}$ and the cutoff scale $\Lambda$. 
This result shows that some of the parameter space of the model may lead to a visible signal in near-future GW experiments, 
allowing us to probe parameter regions that are currently unexplored by other experiments, as discussed below. 
In addition, we note that the relation between the physical parameters of the models and the GW amplitude is given by $\srel\propto \mrel \times \left(\OGW^\text{peak} \right)^{1/12}\,,$ showing a rather mild dependence on the actual amplitude. 

For convenience, the range of relaxion masses, \mrel, and mixing angles with the Higgs boson, \srel, which can be probed via current or future GW experiments as well as the corresponding constraints on the parameter space are summarized in \cref{fig:RelaxionGW}. 
The green and blue/turquoise coloured regions can be accessed with \muAres and the \experiment{Square-Kilometre Array} (\experiment{SKA}) observatory, respectively, depending on the temperature of barrier reappearance.
In the purple region, the reappearance temperature is restricted by current data from the \experiment{North American Nanohertz Observatory for 
  Gravitational Waves} (\experiment{NANOGrav}).
In addition to that, in the grey shaded region, we present the parameter range in which our model can account for the potential GW signal recently observed in \experiment{NANOGrav} data.
The grey dashed line encloses the region in which the relaxion can be trapped without dark photon friction. 
On the other hand, as the figure illustrates, there is a large fraction of parameter space where an additional source of friction is required for the viability of the relaxion mechanism and thus motivates us to add a relaxion-dark photon coupling to the model.
A more detailed discussion of the figure is deferred to \cref{sec:discussion}.

This paper is organized as follows.
We briefly review the relaxion and the dark photon dynamics in \cref{sec:relaxion}.
\Cref{sec:darkphoton,sec:darkmatter} contain a brief discussion of the interplay of the relaxion-dark photon dynamics and the possibility of the relaxion being DM, respectively.
In \cref{sec:constraints} we review the constraints on our model and discuss the available parameter space.  
Subsequently, the production mechanism of the GW background is studied in \cref{sec:GWs}.
We derive the GW spectrum in \cref{sec:GWproduction} and briefly describe how the detectability of the signal is evaluated in \cref{sec:GWdetection}.
The results of this paper are then discussed in \cref{sec:discussion}.
\Cref{sec:conclusion} concludes the paper. 
A brief discussion of the minimal relaxion scenario (without a dark photon coupling) is deferred to \cref{app:minimalScenario}. 
Further details regarding the calculation of the dark photon and GW spectra are provided in \cref{app:DPspectrum,app:GWcalculation}, respectively.

\section{Setup}
\label{sec:relaxion}

In this work, we consider the relaxion $\phi$ coupled to a dark photon field $X_\mu$,
\begin{equation}
	-\mathcal{L} \supset V(H,\phi) + \frac{\rX}{4}\, \frac{\phi}{\frel}\, X_{\mu\nu} \widetilde{X}^{\mu\nu}\,,
\end{equation}
with the potential of the relaxion field~$\phi$ and Higgs doublet~$H$ given by
\begin{equation}
	V(H,\phi) = V_\text{roll}(\phi) + \mu_H^2(\phi) |H|^2 + \lambda |H|^4 + V_\text{br}(H,\phi)\,,
\end{equation}
where $\lambda$ is the Higgs quartic coupling and 
\begin{subequations}
\begin{align}
	V_\text{roll}(\phi) &= - c g \Lambda^3 \phi \,,\\
	\mu_H^2(\phi) &= \Lambda^2 - g \Lambda \phi \,, \\
	V_\text{br}(H,\phi) &= -\frac{\Lbr^4}{\vEW^2} |H|^2 \cos\frac{\phi}{\frel} \,.
\end{align}
\end{subequations}
Here, $c$ is an $\mathcal{O}(1)$ number, $g$ is a dimensionless parameter, $\Lambda$ is the Higgs mass cutoff scale, \Lbr is the backreaction scale, $\vEW = \langle |H| \rangle = \SI{174}{\GeV}$ is the Higgs vacuum expectation value, and \frel is the decay constant of the relaxion. 

During inflation, the relaxion rolls down the linear slope of its potential $V_\text{roll}$. 
It thereby scans the Higgs mass parameter $\mu^2(\phi)$.
Once $\mu^2$ crosses zero, the Higgs acquires a nonvanishing vacuum expectation value, triggering the breaking of the EW gauge symmetry.
The Higgs then backreacts creating wiggles in the relaxion potential via $V_\text{br}$.
Once the Higgs backreaction balances the rolling potential, the relaxion is trapped in the first minimum it encounters.
Choosing $c g \Lambda^3 \frel \sim \Lbr^4$, we end up with a weak-scale expectation value for the Higgs boson, solving the hierarchy problem. 
The relaxion mass and the relaxion-Higgs mixing angle can then be written as~\cite{Banerjee:2020kww,Flacke:2016szy}
\bea
\mrel^2 \simeq \frac{\Lambda_{\rm br}^6}{\frel^2\Lambda\vEW}\,,\quad
	\srel \simeq \sqrt{2} \left(\frac{\mrel^4 \frel \Lambda^2}{\vEW\, m_h^6}\right)^\frac{1}{3} \!,
	\label{eq:massandmixing}
\eea 
 in terms of the theory parameters. 
 Here, $m_h = \SI{125}{\GeV}$ is the Higgs mass.
 
\subsection{Relaxion and dark photon evolution}
\label{sec:darkphoton}

After reheating, the EW symmetry will be restored due to thermal corrections to the potential,  provided that the reheating temperature is above the EW phase transition temperature. 
As a consequence, the relaxion will start rolling again, leading to exponential production of dark photon modes. 
To see the interplay, the coupled differential equations describing the evolution of the spatially homogeneous relaxion and the dark photon modes are given by
 \begin{align}
 &\theta'' + 2 a H \theta' + \frac{a^2}{\frel^2} \frac{\partial V_{\rm roll}}{\partial \theta} = - \frac{a^2}{\frel^2}\, \frac{\rX}{4\,a^4} \left<X_{\mu\nu}\tilde{X}^{\mu\nu}\right> , 
 \label{eq:EOMrelaxion}\\
 &X''_\lambda(k,\tau) + (k^2 - \lambda\, k\, \rX \theta' ) X_\lambda(k,\tau) = 0 \,, 
 \label{eq:EOMdarkphoton}
 \end{align} 
 where $\theta = \phi/\frel$ and primes denote derivatives with respect to conformal time $\tau$ with $a\,d\tau = dt$. 
 We have written the dark photon in terms of its Fourier modes $X_\lambda(\vect{k},\tau)$ in Coulomb gauge, $\nabla\cdot\vect{X}=0$, as
 \begin{equation}
 \hat{\vect{X}}(\vect{x},\tau) = \int\!\frac{d^3 k}{(2\pi)^3} \sum\limits_{\lambda=\pm} X_\lambda(k,\tau) \vect{\varepsilon}_\lambda(\vect{k}) \hat{a}_\lambda(\vect{k}) + \hc 
 \end{equation}
 Here, $\lambda$ denotes the dark photon helicity, and $\vect{\varepsilon}_\lambda$ are the corresponding circular polarization vectors. 

As evident from \cref{eq:EOMdarkphoton}, dark photon modes with $0 < k < \lambda \rX \theta'$ are tachyonic and will experience exponential growth compared to the vacuum fluctuations.
The resulting dark photon spectrum then features anisotropies in its energy-momentum tensor which will act as a source for GW production, leading to a stochastic GW background~\cite{Easther:2006gt,Dufaux:2008dn,Adshead:2018doq,Machado:2018nqk,Machado:2019xuc,Ratzinger:2020oct}. 
Furthermore, since only modes of the helicity with the same sign as the relaxion velocity can become tachyonic, the rolling relaxion will produce a circular polarized dark photon background.
In our case, as we choose $\theta'>0$, only the positive-helicity modes are exponentially produced. 
For these modes, the solution to the equations of motion is given in the Wentzel–Kramers–Brillouin~(WKB) approximation by
\begin{equation}
	\label{eq:WKB}
	X_+(k,\tau) = \frac{e^{g(k,\tau)}}{\sqrt{2\, \Omega(k,\tau)}} \,,
\end{equation}
where 
$\Omega^2(k,\tau) = k\, \rX |\theta'(\tau)| - k^2 > 0$ is the corresponding tachyonic frequency and $g(k,\tau) = \int^\tau\!\!d\tau'\,\Omega(k,\tau')$.
The approximation holds for $|\Omega'/\Omega^2| \ll 1$.

At early times, just after reheating, the friction from dark photons can be neglected. 
Assuming radiation domination, the relaxion velocity can then be written as
\begin{equation}
	\label{eq:thetaBefore}
	\theta'(\tau) = \frac{\Lambda_{\rm br}^4}{5 \frel^2} \left(\frac{\arh}{\tarh}\right)^2\! \tau^3 \left[ 1 - \left(\frac{\tarh}{\tau}\right)^5 \right]\,,
\end{equation}
where  we imposed $\theta'(\tarh) = 0$.
Subsequently, due to the  exponential production, the dark photon friction becomes comparable to the other terms in the equation of motion, \cref{eq:EOMrelaxion}, and asymptotes to the slope of the rolling potential.
The timescale \tapp at which particle production kicks in can be determined by 
\begin{equation}
	\label{eq:XXtilde}
	\frac{\langle X_{\mu\nu}\tilde{X}^{\mu\nu}\rangle(\tau_{\rm pp})}{4 a^4(\tau_{\rm pp})} 
	\approx \frac{\kt^4 e^{2 g(\kt,\tat)}}{4 \pi^2 a^4}
	\sim \frac{\Lbr^4}{\rX}\,,
\end{equation}
where \kt is the mode that dominates the $\langle X \tilde{X} \rangle$ term, given by the saddle point
$\left.\partial g(k,\tau)/\partial k \right|_{\kt}= 0$.	
After \tapp, due to the balance between the potential slope and the backreaction from the dark photon, the relaxion field velocity becomes proportional to the Hubble rate and evolves as~\cite{Choi:2016kke,Banerjee:2018xmn}
\begin{equation}
\label{eq:thetaAfter}
\theta'(\tau) \approx \frac{\xi}{\rX} \,a(\tau) H(\tau) \left( 1 + \epsilon \log \frac{\tau}{\tapp}\right) \approx \frac{\xi}{\rX \tau} \,,
\end{equation}
with a small logarithmic correction ($\epsilon \!\ll\! 1$). 
We defined here the parameter $\xi = \frac{\rX |\theta'|}{a H}$ at \tapp. 
From \cref{eq:XXtilde} we then obtain $\xi \sim \mathcal{O}(10-100)$ with a mere logarithmic dependence on the relaxion parameters~\cite{Choi:2016kke}.
The dominating $k$-mode at each epoch then becomes $\kt/a\sim \rX\dot{\theta}(t)\sim \xi H$.

Once the Universe has cooled sufficiently, the EW phase transition occurs, and the wiggles of the backreaction potential reappear. 
The rolling of the relaxion between reheating and $\Tra$ leads to  a displacement from the minimum in which it originally settled during inflation by
\begin{equation}
	\label{eq:DeltaTheta}
	\Delta\theta = \int\limits_{\tarh}^{\tara} \!d\tau\,\theta' \approx \frac{\xi}{4\,\rX} \left[ 1 + \log\frac{\Hpp^2}{\Hra^2}\right].
\end{equation}  
For the relaxion to remain trapped in this minimum, we need to require that the displacement is less than the distance between the minimum and the next maximum, $\Delta\theta \lesssim 2 \delta$, where $\delta=\Lbr^2/(\Lambda \vEW)$~\cite{Banerjee:2018xmn,Banerjee:2020kww}.
This sets a lower bound on the coupling to the dark photons.  
For smaller couplings, the dark photon friction is insufficient to prevent the relaxion from rolling into one of the neighbouring minima.
The relaxion then needs to traverse $\Delta\theta\sim \mathcal{O}(n)$ to end up in the $n$-th minimum, where $n=1$ denotes the minimum in which it stopped during inflation, extending the parameter space of the theory.
However, going beyond the first minimum requires a careful adjustment of the initial conditions
to let the relaxion stop exactly at the bottom of the $n$-th minimum at reappearance.
Otherwise, the time required for the relaxion to reach the bottom would exceed the age of the Universe.
We thus simply assume $\rX = \xi/(2\delta)$ in the following.\footnote{
	As $\delta = \frac{\Lbr^2}{\Lambda\vEW} < \frac{\vEW}{\Lambda}$, this implies $\rX \gtrsim 10^2 \left(\frac{\Lambda}{\SI{1}{\TeV}}\right)\left(\frac{\xi}{\num{10}}\right)$. 
	Such large couplings can be obtained in a technically natural way for example via the clockwork mechanism~\cite{Choi:2014rja,Choi:2015fiu,Kaplan:2015fuy,Giudice:2016yja}.  
}
 
\subsection{Relaxion dark matter}
\label{sec:darkmatter}

After the reappearance of the Higgs backreaction potential, the displaced relaxion begins to oscillate around the minimum of its potential, providing a candidate for ultralight DM as discussed in Ref.~\cite{Banerjee:2018xmn}. 
Assuming the maximal displacement of $\Delta\theta = 2 \delta$, the relaxion relic abundance is given by
\begin{equation}
	\Omega_\phi = \frac{2 \,\mrel^2 \frel^2 \delta^2}{3 \Mpl^2 H_0^2} \frac{g_{*s}(T_0)T_0^3}{g_{*s}(\Tosc)\Tosc^3}\,,
\end{equation}
where $\Tosc\sim \min[\Tra,\sqrt{m_\phi \Mpl}]$ is the temperature when the relaxion starts to oscillate. 
Requiring that the relaxion provides all of DM, i.e.,~$\Omega_\phi h^2=0.12$~\cite{Aghanim:2018eyx}, the relaxion decay constant can be expressed as
\bea
	\label{eq:fDM}
	\!\!\! \frel \sim \SI{5e9}{\GeV} \left(\frac{\Lambda}{\SI{1}{\TeV}}\right)^{\!\frac{2}{5}} \!\left(\frac{ \Tosc}{\SI{1}{\GeV}} \right)^{\!\frac{9}{10}} \!\left(\frac{\SI{0.1}{\eV}}{m_\phi}\right)\!.\!\!\!
\eea

As we are considering coherently oscillating relaxion DM here, its mass needs to be less than approximately \SI{10}{\eV} in order to be described by a classical field~\cite{Kolb:1990vq}.
For this range of relaxion masses, the possible decay channels are into two photons and two dark photons, $\Gamma_\phi=\Gamma_{\gamma\gamma}+ \Gamma_{XX}$.
The decay rate into a dark photon pair is given by
\begin{equation}
	\Gamma_{XX} = \frac{\rX^2}{64\pi}\frac{\mrel^3}{\frel^2} \,,
\end{equation}
while the decay width of $\phi\to\gamma\gamma$ is subdominant compared to that of $\phi\to X X$ as it is suppressed by the square of the relaxion-Higgs mixing angle~\cite{Banerjee:2018xmn}, which in turn is bounded from above by $\srel \lesssim v_H/\frel$~\cite{Flacke:2016szy}. 
The relaxion lifetime hence becomes
\begin{equation}
	\tau_\phi \sim \SI{20}{\giga\year} \left(\frac{25}{\xi}\right)^{2} \left(\frac{\Tosc}{\SI{1}{\GeV}}\right)^{3} \left(\frac{\SI{0.1}{\eV}}{\mrel}\right)^5,
\end{equation}
where we have chosen 
$\rX = \xi/(2\delta)$. 
Since the decay of DM into relativistic particles
affects the spectrum of the cosmic microwave background~(CMB) at
low-$\ell$ multipoles, the lifetime is constrained as $\tau_\phi>\SI{160}{\giga\year}$~\cite{Audren:2014bca}.

As shown in Refs.~\cite{Agrawal:2017eqm,Kitajima:2017peg,Kitajima:2020rpm,Ratzinger:2020oct}, 
for $\rX(\Delta\theta)_{\rm sep}\sim\mathcal{O}(10^2)$, an oscillating ALP may introduce a second phase of tachyonic dark photon production, which could suppress the DM abundance by up to two orders of magnitude. This condition is satisfied in parts of the parameter space where the field displacement is maximal, since there $\rX\sim\mathcal{O}(\xi/\delta)$ and $\xi\sim \mathcal{O}(10-10^2)$, and introduces some uncertainty in our estimate of the DM abundance in those regions.

\subsection{Constraints}
\label{sec:constraints}

A successful cosmological relaxation of the Higgs mass requires 
the inflation sector to dominate the total energy density, $3 H_\text{I}^2\Mpl^2\gtrsim \Lambda^4 $, as well as that the classical motion of the relaxion dominates over quantum fluctuation during inflation, $(\Delta\phi)_{\rm cl}\gtrsim H_{\rm I}/2 \pi$.
Here, $H_\text{I}$ is the Hubble scale during inflation.
Combining these two constraints, we get an upper bound on the cutoff scale $\Lambda$,
\bea
\Lambda\lesssim \left(\frac{ 2\pi\sqrt{3}\, \Mpl^3\Lbr^4}{f}\right)^{1/6}\!\!.
\label{eq:constraintslambdacosmo}
\eea 
As we are considering a Higgs-dependent backreaction potential, we also require $\Lbr\lesssim \vEW$~\cite{Graham:2015cka,Flacke:2016szy}.\footnote{See Refs.~\cite{Hook:2016mqo,Fonseca:2019lmc} for a discussion of other stopping mechanisms in which this constraint can be relaxed.}
The allowed range of the effective cutoff $\Lambda$ of the theory is
\bea
	\Mpl \gtrsim \frel \gtrsim \Lambda \gtrsim 4 \pi \vEW \sim \SI{1}{\TeV}\,.
	\label{eq:constraintsf}
\eea 
Also, for a generic backreaction potential which does not change the late-time cosmology, the range of reappearance temperatures is 
\bea
\vEW\gtrsim \Tra\gtrsim T_{\rm BBN}\sim \SI{10}{\MeV}\,.
\label{eq:constraintslTra}
\eea

For masses below the \si{\eV} scale, the relaxion can further mediate long-range forces.  
Experiments looking for such interactions (fifth force experiments, inverse-square-law, and equivalence-principle tests) constrain the coupling of the relaxion to ordinary matter~\cite{OHare:2020wah,Tan:2020vpf,Berge:2017ovy,Chen:2014oda,Kapner:2006si,Schlamminger:2007ht,Hoskins:1985tn}, which is induced by the relaxion-Higgs mixing angle given in~\cref{eq:massandmixing}.
In a similar manner, for masses up to the \si{\keV} range, the mixing is constrained from stellar cooling~\cite{Grifols:1988fv,Cadamuro:2011fd,Raffelt:2012sp,Hardy:2016kme}, as it induces relaxion-mediated contributions to the energy loss in stars.
Slightly weaker limits on the mixing angle can furthermore be obtained from bounds on the solar relaxion flux as constrained by \experiment{XENON1T} and other dark-matter direct detection experiments~\cite{Budnik:2019olh}.

Additional constraints arise when coupling the relaxion to a dark photon field (see \cref{app:minimalScenario} for a discussion of the minimal scenario without this coupling), with the coupling here chosen to saturate the trapping condition, i.e., $\rX = \xi/(2\delta)$. 
For the dark photon induced friction to trap the relaxion, reappearance has to occur sufficiently late for the dark photon to be produced, i.e., $\Hpp > \Hra$. 
This sets a lower bound on the relaxion mass, $\mrel\gtrsim \sqrt{10}\,\delta \Hra$, for which the dark photon scenario can be applied.

If this condition is satisfied, throughout its evolution from reheating to reappearance, the relaxion continuously produces dark photons, depositing energy density into the latter.  
At the time of trapping, $t_{\rm ra}$, the dark photon energy-density can be estimated as
\begin{equation}
	\label{eq:rhoX}
	\rho_X(t_{\rm ra}) = \frac{1}{a^4(t_{\rm ra})} \int^{t_{\rm ra}} \!\! dt^\prime a^4(t^\prime)\dot{V} = \frac{1}{2}\, \mrel^2 \frel^2 \,.
\end{equation} 
For the Universe to remain radiation dominated, we require $\rho_X(\Tra)$ to be smaller than the radiation energy density at $\Tra$, i.e.\ $\rho_X(\Tra)\lesssim 3\Mpl^2\Hra^2\sim \Tra^4$; otherwise, the relaxion dominates the total energy density of the Universe.
After $\Tra$, dark photon production stops, and the corresponding energy density redshifts as that of radiation.  

The dark photon is further constrained by its contribution to the energy density of the Universe at low temperatures, typically parametrized in terms of the effective number of neutrino species \Neff.
For temperatures $T\lesssim m_e$, after electrons and positrons have annihilated, the dark photon energy density (or of any additional relativistic species for that matter) can be written in terms of its contribution $\Delta\Neff$ to \Neff as
\begin{equation}
	\label{eq:Neff}
	\rho_X(T) = \frac{7\,\pi^2}{120} \Delta\Neff \left(\frac{4}{11}\right)^\frac{4}{3} T^4 ,
\end{equation}
where the $4/11$ factor accounts for the difference between the neutrino and photon temperature. 
Using \cref{eq:rhoX} and the current \SI{95}{\%}~C.L.\ limit $\Neff = 3.27\pm0.29$ from CMB data~\cite{Aghanim:2018eyx} combined with local measurements of the Hubble rate~\cite{Riess:2018uxu}, as well as $\Neff^\text{SM} = 3.046$ in the Standard Model, we obtain a lower limit on the reappearance temperature as a function of the relaxion mass and decay constant
\bea
\Tra \gtrsim  
2.5\, g_{*s,\ra}^{-1/3} \, \sqrt{\mrel\frel}\,,
\label{eq:minimumTra}
\eea
where $g_{*s,\ra}$ is the number of entropic degrees of freedom at reappearance. 

If we assume that 
the relaxion accounts for the full DM abundance, then plugging \cref{eq:fDM} to \cref{eq:minimumTra}, we get, 
\begin{equation}
	\Tra \gtrsim 
	\SI{450}{\MeV}\left(\frac{67}{g_{ *s,\ra}}\right)^{\frac{1}{3}} \left(\frac{\Lambda}{\SI{10}{\TeV}}\right)^{\!\frac{4}{11}} .
	\label{eq:DPNeff}
\end{equation}
Here, we also assume that the relaxion starts to oscillate at \Tra, which is only true for a sufficiently heavy relaxion.
As we require $\Lambda  \gtrsim 4 \pi \vEW \sim \SI{1}{\TeV}$, this sets a lower bound of $\Tra \gtrsim \SI{240}{\MeV}$ on the reappearance temperature for the relaxion DM scenario to be realized.\footnote{Note that, for this value of the reappearance temperature, the relaxion starts to oscillate directly after reappearance as long as $\mrel \gtrsim \SI{5e-10}{\eV}$.}
Upon the same assumptions, $\Lambda_{\rm br}\lesssim\vEW$ further sets an upper bound on the reappearance temperature, 
\bea
\! \Tra \lesssim \min\left[\vEW,\ \SI{80}{\GeV}\, \left(\frac{\SI{e6}{\GeV}}{\Lambda}\right) \left(\frac{96}{g_{\rm *s,ra}}\right)^{\!\!\frac{1}{3}} \right].
\label{eq:Tra_dm_ub} 
\eea 

\begin{figure}
	\includegraphics[width=\columnwidth]{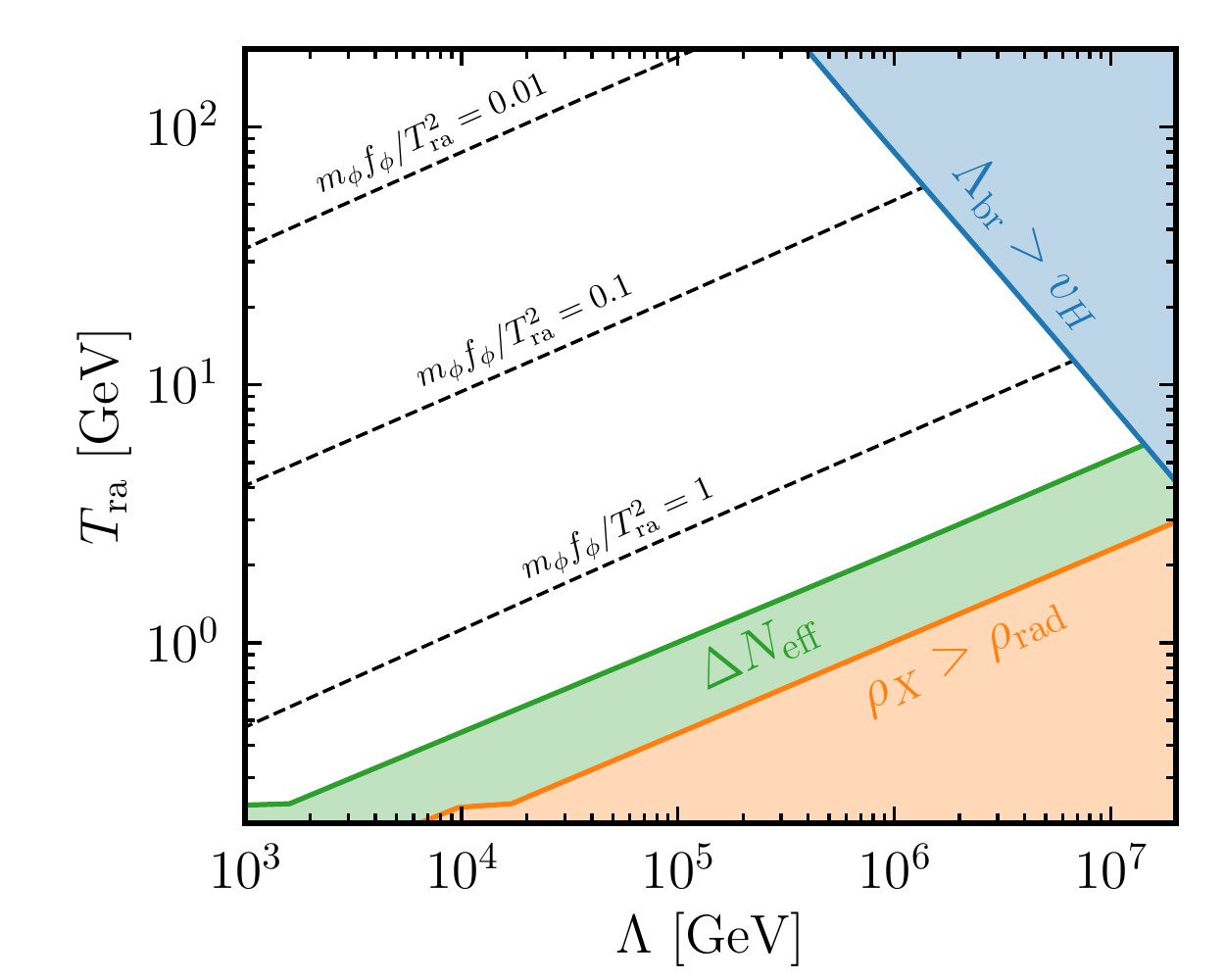}
	\caption{%
		Allowed range of reappearance temperatures \Tra as a function of the cutoff scale $\Lambda$, while fixing \frel to the value reproducing the measured DM abundance, \cref{eq:fDM}, assuming $\Tosc = \Tra$.
		The blue shaded region is excluded since $\Lbr > \vEW$, while in the green shaded regions, the CMB bound on $\Delta\Neff$, \cref{eq:DPNeff}, is violated. 
		In the orange shaded region, the dark photon energy density further dominates the Universe at reappearance.
		The dashed lines are contours of constant $\mrel \frel / \Tra^2$, which sets the amplitude of the GW spectrum; cf.~\cref{eq:AmplitudeEstimate}. 
	}
	\label{fig:Tra}
\end{figure}

In \cref{fig:Tra}, we show the minimal and maximal allowed reappearance temperature for relaxion DM as a function of the cutoff $\Lambda$ of the theory.
Combining $\Lbr < \vEW$~(blue) and the \Neff constraint~(green), we see that the highest $\Lambda$ for which the relaxion can be realized as coherently oscillating DM is $\Lambda \lesssim \SI{e7}{\GeV}$, which is in accordance with the constraints for $\Tra \simeq \SI{6}{\GeV}$.
Because of the rapid change in the radiative degrees of freedom around the time of the QCD phase transition, the $\Delta\Neff$ limit on the reappearance temperature saturates at $\Tra \sim T_\text{QCD}$ for $\Lambda \lesssim \SI{2}{\TeV}$.
We also depict the weaker bound $\rho_X < \rho_\text{rad}$ from the total energy density in orange.

\begin{figure*}
	\centering
	\includegraphics[width=2\columnwidth]{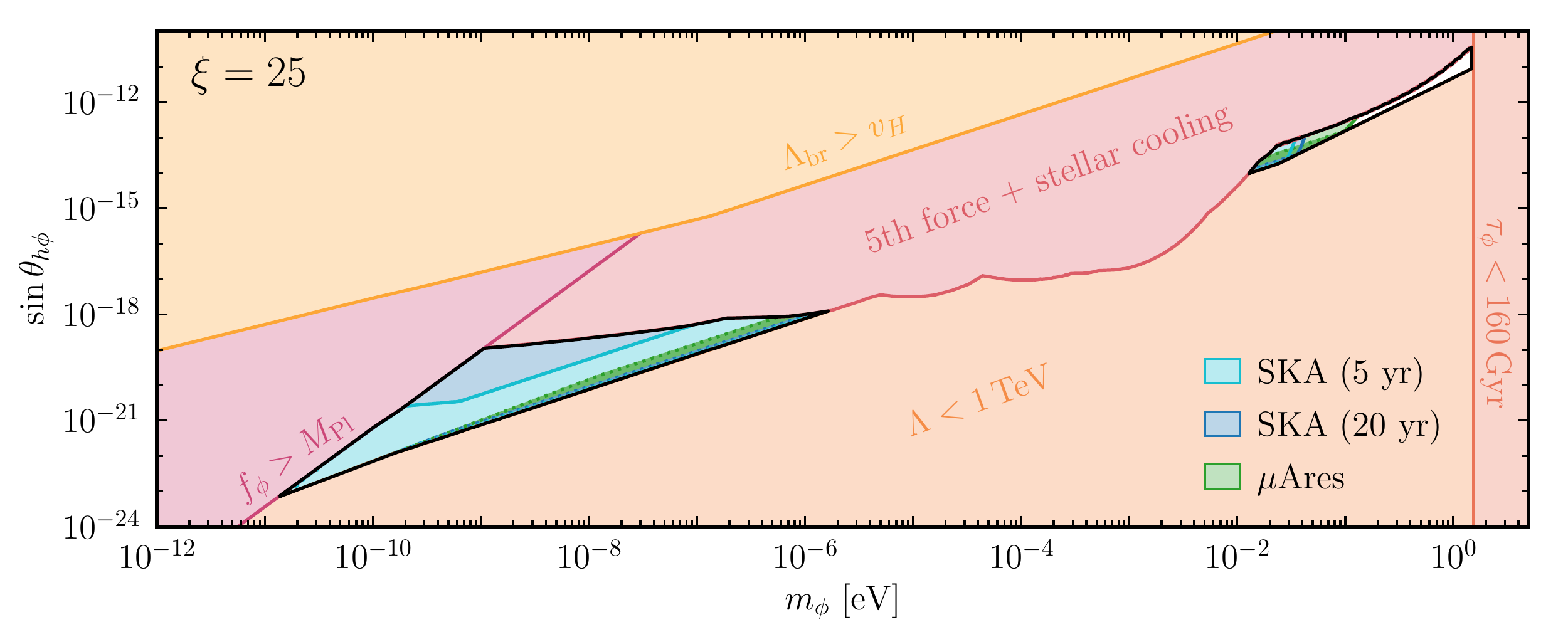}
	\caption{%
		Available parameter space (black framed region) for relaxion DM in the relaxion mass \mrel vs.\ mixing angle \srel plane.
		The red and orange shaded regions are excluded by the indicated constraints of combinations thereof.	
		The colored regions inside the viable DM space can be probed via GWs in \muAres~(green) or \experiment{SKA}~(blue/turquoise).
		The light shading and solid lines indicate points that can be probed for a subrange of reappearance temperatures, whereas the darker shaded parts enclosed by dotted lines are accessible for all valid \Tra.
		An animated version of the plot scanning the reappearance temperature is enclosed in the ancillary material of this work.
	}
	\label{fig:DMGW}
\end{figure*}

In \cref{fig:DMGW}, we depict the constraints on the relaxion parameters as a function of the relaxion mass \mrel and the mixing angle \srel, where we determined $\Lambda$ and \frel from the DM abundance using \cref{eq:fDM,eq:massandmixing}.
The red and orange shaded regions are excluded by the indicated constraints,
where the shape of the exclusions now partially differs from the corresponding ones in \cref{fig:RelaxionGW} as we require here $\Omega_\phi = \Omega_\text{DM}$. 
Scanning over all allowed values of \Tra, the full range of masses and mixing angles for which we can obtain coherently oscillating relaxion DM is indicated by the black-framed regions in \cref{fig:DMGW}.
We obtain two separated islands of viable parameter space, one at low masses, $\SI{e-11}{\eV} \lesssim \mrel \lesssim \SI{e-6}{\eV}$, with mixing angles around $\num{e-23} \lesssim \srel \lesssim \num{e-18}$, and another island at high masses, $\SI{e-2}{\eV} \lesssim \mrel \lesssim \SI{1}{\eV}$, with a narrow range of mixing angles around $\num{e-14} \lesssim \srel \lesssim \num{e-11}$.
Note that, in the high-mass island, the coupling to dark photons is required to trap the relaxion, whereas in most of the low-mass island, relaxion DM can be realized without dark photon friction~\cite{Banerjee:2018xmn}.
It shall, moreover, be emphasized that the low-mass DM regions in the minimal and dark photon scenario are separated in the reappearance temperature, since relaxion stopping via Hubble friction requires $\mrel \lesssim \sqrt{8}\, \delta \Hra$, whereas dark photon production only occurs for $\mrel \gtrsim \sqrt{10}\, \delta \Hra$. 
See \cref{app:minimalScenario} for a detailed discussion on the minimal scenario.

\section{Gravitational waves}
\label{sec:GWs}

Let us now consider the stochastic GW background generated from anisotropies in the energy-momentum tensor of the dark photon produced during the postinflationary evolution of the relaxion.
In particular, we focus here on the GWs sourced during the rolling of the relaxion between reheating and the EW phase transition.
Further gravitational radiation can be generated when the relaxion oscillates around the minimum of its potential, as explored in Refs.~\cite{Machado:2018nqk,Machado:2019xuc,Ratzinger:2020oct,Salehian:2020dsf,Soda:2017dsu,Kitajima:2018zco,Chatrchyan:2020pzh} in the context of general ALP models, or in a confining phase transition that generates the backreaction potential~\cite{Barducci:2020axp}.
In combination, the various sources may lead to an interesting and rich GW spectrum with multiple peaks.

The spectrum of a stochastic background of GWs is characterized by its fractional energy density, i.e.~the energy density normalized to the critical value, $\rhoc = 3 \Mpl^2 H_0^2$, per logarithmic frequency interval,
\begin{equation}
	\OGW(\freq) = \frac{1}{\rhoc} \frac{d \rhoGW}{d \log\freq} \,,
\end{equation}
where the total GW energy density is given by
\begin{equation}
	\label{eq:rhoGWtot}
	\rhoGW = \frac{\Mpl^2}{4} \left<\dot{h}_{ij} \dot{h}^{ij}\right>
	= \int\!\frac{d\freq}{\freq}\frac{d \rhoGW}{d \log \freq}\,.
\end{equation}
Here, $h_{ij}$ denotes the GW metric perturbations, and the dot indicates the derivative with respect to cosmic time~$t$.
Switching to conformal time $\tau$, $dt = a\,d\tau$, where $a$ is the scale factor of the Universe, the metric reads
\begin{equation}
	d s^2 = a(\tau) \left\{ d\tau^2 - \right[\delta_{ij} + h_{ij}(\vect{x},\tau)\,\left] d x^i d x^j \right\} \,.
\end{equation}
During radiation domination, the Einstein equations in the linear regime for the metric perturbations in momentum space using transverse-traceless gauge become
\begin{equation}
	( \partial_\tau^2 + k^2 )\, a(\tau)\, h_{ij}(\vect{k},\tau) = \frac{2\,a(\tau)}{\Mpl^2} \Pi_{ij}(\vect{k},\tau)\,,
\end{equation}
where $k=|\vect{k}|$ is the comoving wave number. 
The anisotropic stress tensor $\Pi_{ij}$ relates to the energy-momentum tensor $T_{ij}$, via
$\Pi_{ij}(\vect{k},\tau) = \Lambda_{ij}^{ab}(\vect{k}) T_{ab}(\vect{k},\tau)$,
where $\Lambda^{ab}_{ij} = P_i^{\,a} P_j^{\,b} - \frac{1}{2} P_{ij} P^{ab}$ with $P_{ij} = \delta_{ij} - k_i k_j/k^2$ being the projector that extracts the transverse and traceless part~\cite{Caprini:2018mtu}. 
The equations of motion are then solved by (neglecting the $a^{\prime\prime}$ term which vanishes in a radiation-dominated universe i.e. for $a\propto \tau$)
\begin{equation}
	\label{eq:hij}
	\hat{h}_{ij}(\vect{k},\tau) = \frac{2}{\Mpl^2}\,\int^\tau\!\!\!d\tau' \,\frac{a(\tau')}{a(\tau)}\, \hat{\Pi}_{ij}(\vect{k},\tau)\,\mathcal{G}(k,\tau,\tau')\,,
\end{equation}
where $\mathcal{G}(k,\tau,\tau')=\sin[k(\tau-\tau')]/k$ is the causal Green's function. 
We denote here the operator form of any quantity $Q$ by $\hat{Q}$.   

\subsection{Gravitational wave production}
\label{sec:GWproduction}

The GW energy density per logarithmic interval in the comoving momentum $k$ of a generic stochastic source at conformal time $\tau$ is given by~\cite{Caprini:2018mtu}
\begin{multline}
	\frac{d \rhoGW}{d \log k}(k,\tau) = 
	\frac{k^3}{4 \pi^2 \Mpl^2 a^4(\tau)}
	\int\limits_{\tau_i}^{\tau}\!\!d\tau' \!\!\int\limits_{\tau_i}^{\tau}\!\!d\tau' 
	a(\tau')\, a(\tau'')\, 
	\times\\
	\cos[k(\tau'-\tau'')]\,
	\Pi^2(k,\tau',\tau'') \,,
\end{multline}
where $\tau_i$ is the time at which the GW source starts operating and $\Pi^2(k,\tau',\tau'')$ is the unequal time correlator of the anisotropic stress, defined as
$\langle0| \hat{\Pi}^{ab}(\vect{k},\tau) \hat{\Pi}^{\ast}_{ab}(\vect{k}',\tau')|0 \rangle = (2\pi)^3 \delta(\vect{k}-\vect{k}') \Pi^2(k,\tau,\tau')$.
In our case, the GWs are generated between reheating and reappearance; hence $\tau_i = \tarh$ and $\tau \leq \tara$. 
As the GWs produced before the relaxion reaches its terminal velocity will, however, be subdominant, we can take $\tau_i = \tapp$, so to first approximation, the GW signature becomes independent of the temperature to which the Universe was reheated.

The dark photon anisotropic stress sourcing the GWs can be written in terms of the dark electric and magnetic fields as
\begin{multline}
	\label{eq:Piab}
	\hat{\Pi}_{ab}(\vect{k},\tau) = -\frac{\Lambda_{ab}^{ij}(\vect{k})}{a^2(\tau)} \int \!\frac{d^3 q}{(2\pi)^3}\, \big[ \hat{E}_i(\vect{q},\tau) \hat{E}_j(\vect{k}-\vect{q},\tau) \,+ \\
	\hat{B}_i(\vect{q},\tau) \hat{B}_j(\vect{k}-\vect{q},\tau) \big] \,.
\end{multline}

Focusing on the dominant modes which have completed their phase of maximal tachyonic growth, $q \gtrsim \rX |\theta'|/2$,
we find that GWs with momentum $k$ are dominantly produced at the time of reappearance.
For frequencies below the peak, both dark photon modes in \cref{eq:Piab} have momenta close to the one that experiences maximal growth at reappearance, $|\vect{q}| \sim |\vect{k}-\vect{q}| \sim \rX |\theta'_\ra|/2$, whereas above the peak, one of the contributing modes must have a larger momentum and therefore have exited the tachyonic band before reappearance.
Details of the calculation are deferred to \cref{app:GWcalculation}.
The present-day GW spectrum can then be written as
\begin{equation}
	 \label{eq:GWspectrum}
	\OGW(\freq) = \OGW^\text{peak}\, \mathcal{S}_\xi\left({\freq}/{\fp}\right)
\end{equation}
with the peak frequency
\begin{equation}
	\label{eq:fPeak}
	\fp = \frac{k_\ra}{2\pi\,a_0} = \frac{\rX |\theta'_\ra|}{2\pi\,a_0}
	= \frac{\ara}{a_0} \frac{\xi\,\Hra}{2\pi} 
	\,,
\end{equation}
where $k_\ra$ is the mode that exits the tachyonic band at reappearance, and the peak amplitude
\begin{equation}
	\label{eq:OmegaPeak}
	\!\OGW^\text{peak} = \frac{1}{\rhoc}\,\frac{\ara^4}{a_0^4}\,\frac{1}{\Mpl^2 \Hra^2}\,\frac{\Lbr^8}{4\,\rX^2} 
	= \frac{1}{\rhoc}\,\frac{\ara^4}{a_0^4}\,\frac{\mrel^4 \frel^4}{\Mpl^2 \xi^2 \Hra^2} 
	\,.\!
\end{equation}
We used \cref{eq:thetaAfter} here as well as $\xi \simeq 2 \delta \rX$ and $\Lbr^4=\mrel^2\frel^2/\delta$ to recast \cref{eq:fPeak,eq:OmegaPeak} into the estimations in \cref{eq:peakestimate,eq:AmplitudeEstimate}.
The spectral shape $\mathcal{S}_\xi$ is given by
\begin{equation}
	\label{eq:SpectralShape}
	\mathcal{S}_\xi(x) = \frac{1}{1 + \frac{48}{5} (x-1)^4 + \frac{\num{19965}}{256\,\xi^2} \left(x^{-3} + 3\,x - 4\right)}\,.
\end{equation}
Note that, similar to the oscillating axion case~\cite{Machado:2018nqk,Machado:2019xuc,Salehian:2020dsf,Ratzinger:2020oct}, we obtain a GW spectrum with an unpolarized low-frequency tail and a chiral spectrum above the peak.

\subsection{GW detection}
\label{sec:GWdetection}

Having predicted the GW spectrum generated by the relaxion dynamics, we can now evaluate its detectability in current and future experiments.
A stochastic GW background can be detected in a given experiment if its signal-to-noise ratio~(SNR)~\SNR exceeds a threshold value $\SNR_\text{thr}$. 
The SNR is given by~\cite{Thrane:2013oya}
\begin{equation}
	\label{eq:SNR}
	\SNR^2 = T_\text{obs} \!\!\int\limits_{\freq_\text{min}}^{\freq_\text{max}} \!\!\!d\freq \left[\frac{\OGW(\freq)}{\Omega_n(\freq)}\right]^2 \,,
\end{equation}
where $T_\text{obs}$ is the period of observation, $(\freq_\text{min},\freq_\text{max})$ is the frequency range of the detector, and $\Omega_n(\freq)$ is the detector's noise spectrum converted to fractional energy density.
For a cross-correlated measurement in a network of detectors, as, for instance, a pulsar timing array (PTA), the noise spectrum has to be replaced by the effective noise \Oeff of the network (see Ref.~\cite{Thrane:2013oya} for further details), and the SNR is given by \cref{eq:SNR} with an additional factor of~2.  

We present here the projected sensitivities of the planned space-based \experiment{Laser Interferometer Space Antenna (LISA)}~\cite{Audley:2017drz,Cornish:2018dyw},
the planned \experiment{Square-Kilometre Array (SKA)}~\cite{Janssen:2014dka} PTA observatory,%
\footnote{%
	We assume here that the prospective foreground from supermassive black hole binaries can be subtracted.
	Further details on the sensitivities can be found in Ref.~\cite{Breitbach:2018ddu}.
}
and the proposed microhertz experiment \muAres~\cite{Sesana:2019vho}.
In addition, we evaluate current limits from the  \experiment{NANOGrav} 11~year dataset~\cite{Arzoumanian:2018saf}.
Other GW observatories such as ground-based interferometers or potential \experiment{LISA}-successor experiments in the decihertz regime do not cover the frequency range in which a signal is expected in our scenario.

Recently, \experiment{NANOGrav} has further reported strong evidence for a common-spectrum stochastic process across the pulsars in their 12.5~year dataset~\cite{Arzoumanian:2020vkk}, which might be due to a GW background.
However, a GW detection could not be established yet, due to the lack of conclusive evidence regarding the interpulsar correlations of this process.
Nonetheless, we also present here an interpretation of this signal as the GW background generated in our model, fitting our signal to the \experiment{NANOGrav} data based on the procedure outlined in Ref.~\cite{Ratzinger:2020koh}.

Furthermore, GWs contribute to the total amount of radiation in the Universe and are therefore subject to constraints on \Neff.
This sets an upper bound on the peak amplitude of the spectrum. 
Using \cref{eq:Neff}, we obtain
\begin{equation}
	h^2\OGW^\text{peak} 
	\leq \frac{a_\text{rec}^4}{a_0^4} \frac{7 \pi^2}{120} \left(\frac{4}{11}\right)^\frac{4}{3} \frac{T_\text{rec}^4}{\rhoc/h^2} \frac{\Delta N_\text{eff}}{\mathcal{I}_\mathcal{S}(\xi)}
	\approx \num{e-6},\!
\end{equation}
where $\mathcal{I}_\mathcal{S}(\xi) = \int\! d\log x\,\mathcal{S}_\xi(x) \simeq \numrange{0.8}{1.5}$ is the integral over the spectral shape and $T_\text{rec}$ is the photon temperature at recombination.

An even stronger bound is obtained from the contribution of the dark photons to $\Delta\Neff$. Since the GWs are sourced by dark photon anisotropies, this directly leads to an upper bound on the GW amplitude which is stronger than the one from the direct contribution of the GWs to $\Delta\Neff$. 
Plugging \cref{eq:minimumTra} into \cref{eq:OmegaPeak}, we  obtain
\begin{equation}
	h^2\OGW^\text{peak} \lesssim \num{e-8} \left(\frac{g_{\ast,\ra}}{106.75}\right)^{\!\frac{1}{3}} \left(\frac{25}{\xi}\right)^2 \left(\frac{\Delta \Neff^X}{0.52}\right)^2,
\end{equation}
where the superscript $X$ indicates that this is the dark photon contribution to $\Delta\Neff$. A similar constraint on the GW amplitude was already observed in Ref.~\cite{Ratzinger:2020koh} in the audible axion scenario. There, it was noted that the parameter region that provides the best fit to the stochastic GW background that was potentially observed by the NANOGrav collaboration~\cite{Arzoumanian:2020vkk} is in tension with constraints on $\Delta \Neff$.

\section{Discussion }
\label{sec:discussion}

\begin{figure*}
	\centering
	\includegraphics[width=2\columnwidth]{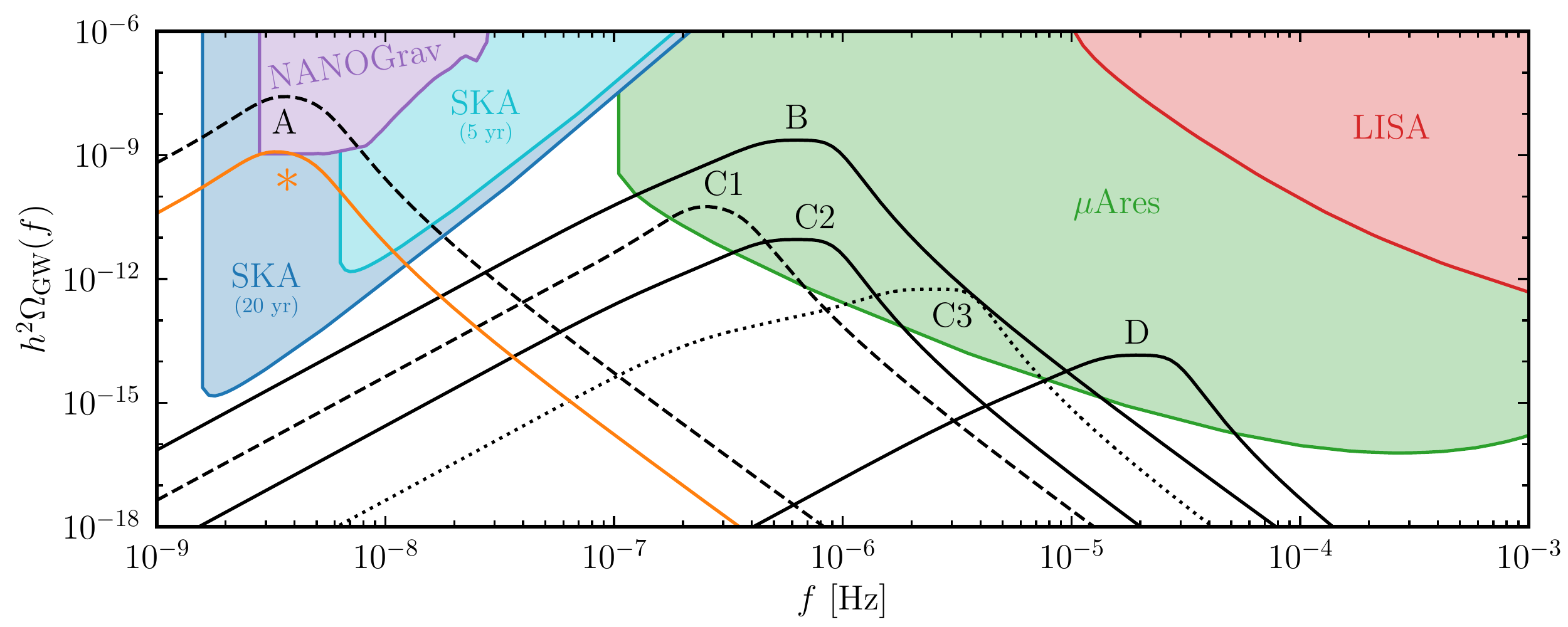}
	\caption{%
		GW spectra for the parameter points listed in \cref{tab:BM}.
		The orange spectrum labeled ``$\ast$'' corresponds to our best fit to the \experiment{NANOGrav} 12.5~year dataset.
		The colored regions depict the projected power-law integrated sensitivity curves of \experiment{LISA}~(red), \muAres~(green), and \experiment{SKA}~(turquoise and blue, assuming 5 and 20~years of observation, respectively) and the  current exclusion from the \experiment{NANOGrav} 11~year data set~(purple).
	}
	\label{fig:spectra}
\end{figure*}

\begin{table}
		\begin{tabular}{lcccc}
		\hline\hline
		\rule[-4pt]{0pt}{14pt}& $\mrel$ (eV) & $\frel$ (GeV) & $\Tra$ (GeV) & $\xi$ \\ \hline
		A\hspace*{2em} & \num{3e-9} & \num{e14} & \num{0.02} & 10 \rule{0pt}{10pt} \\
		B &  \num{2e-3} & \num{e12} & \num{1} & 25 \rule{0pt}{10pt}  \\
		C1 & \num{5} & \num{e8} & \num{1} & 10 \rule{0pt}{10pt} \\
		C2 & \num{5e-3} & \num{e11} & \num{1} & 25 \rule{0pt}{10pt} \\
		C3 & \num{5e-6} & \num{e14} & \num{1} & 100 \rule{0pt}{10pt} \\
		D & \num{0.1} & \num{e12} & \num{30} & 25 \rule{0pt}{10pt} \\
		$\ast$ & \num{e-9} & \num{e14} & \num{0.02} & 10 \rule{0pt}{10pt} \\
		\hline\hline
	\end{tabular}
	\caption{Parameter values for the spectra shown in \cref{fig:spectra}}
	\label{tab:BM}
\end{table}

In \cref{fig:spectra} we show some example GW spectra.
The corresponding parameter values are listed in \cref{tab:BM}.
The coupling to dark photons is set to $\rX = \xi/(2\delta)$.
In addition, the projected sensitivities of \experiment{LISA}~(red), \muAres~(green), and \experiment{SKA} with an observation period of 5~years~(turquoise) and 20~years~(blue), represented by the corresponding power-law integrated sensitivity curves~\cite{Thrane:2013oya}, are depicted as shaded regions.
The current \SI{95}{\%} upper limits from the 11~year \experiment{NANOGrav} data set~\cite{Arzoumanian:2018saf} are shown in purple.
The example spectra B -- D are detectable with \muAres, whereas the benchmark A is accessible via \experiment{SKA} and is excluded by \experiment{NANOGrav}.
The benchmark labeled ``$\ast$'' shown in orange corresponds to our best fit to the \experiment{NANOGrav} 12.5~year dataset as discussed further below.

At low frequencies, the GW spectra behave approximately as $f^3$, in accordance with the expectations based on causality arguments using that the anisotropic stress of a causal source cannot be correlated at scales above the horizon size at the time of production~\cite{Caprini:2009fx,Caprini:2018mtu}.
At high frequencies, the spectra fall approximately like $f^{-4}$, allowing a simple distinction from the much steeper falling GW background generated from oscillating~\cite{Machado:2018nqk,Machado:2019xuc,Salehian:2020dsf,Ratzinger:2020oct} or rotating~\cite{Co:2021rhi} axionlike fields.
It should further be noted that, when the peak position is fixed, changing $\xi$ barely affects the UV tail, while the IR tail goes as $\xi^2$ (cf.\ \cref{fig:spectralShape}), potentially allowing to one disentangle the reappearance temperature and $\xi$ in the peak frequency \cref{eq:fPeak} and thereby facilitating the determination of the relaxion parameters from a hypothetical observed signal.
Larger values of $\xi$ further result in a flatter peak, although this may be an artifact of our analytic approximation; cf.\ \cref{eq:SpectralShape}.

\begin{figure}
	\centering
	\includegraphics[width=\columnwidth]{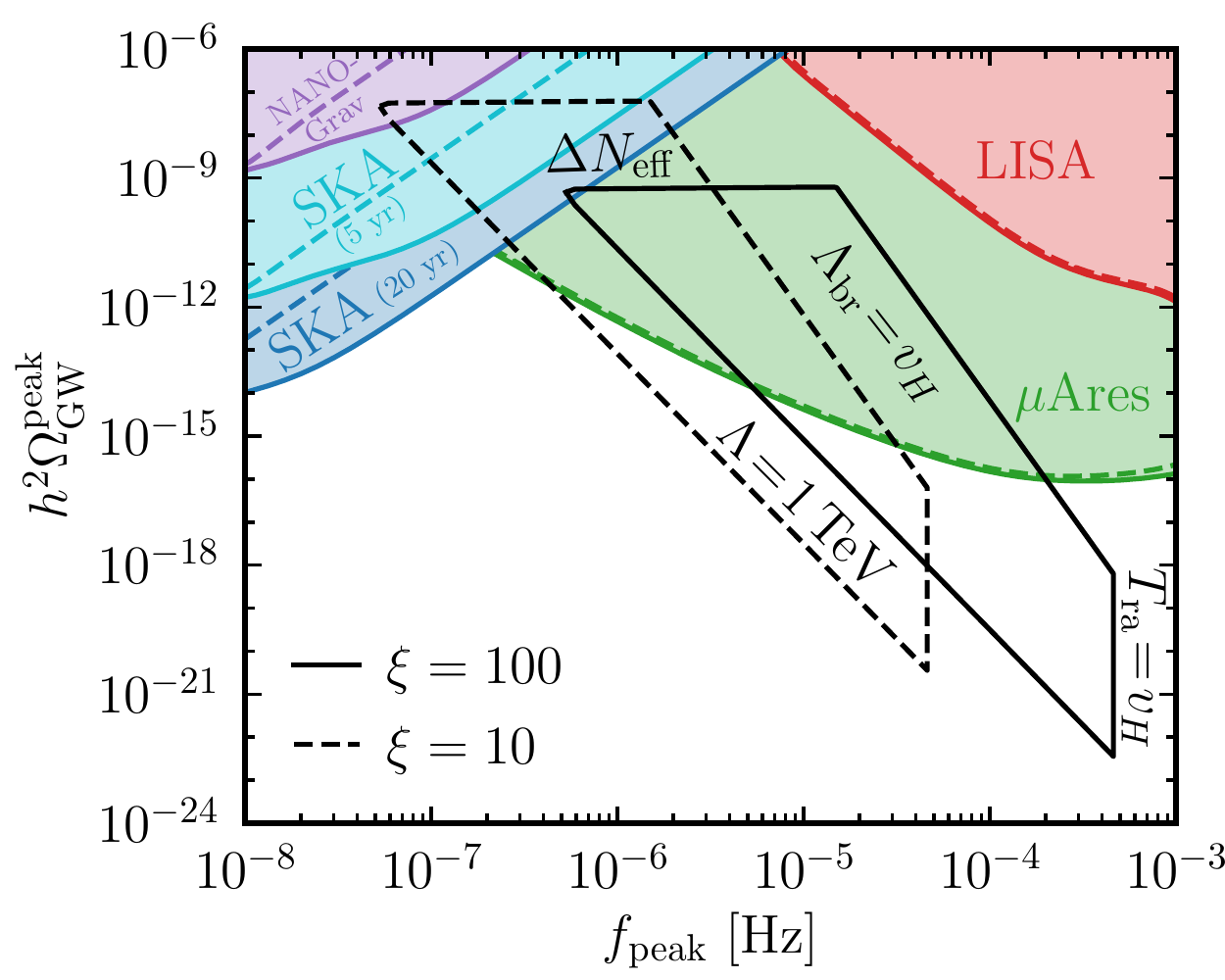}
	\caption{%
		Values of the peak frequency and amplitude of the GW spectrum which can be obtained in the relaxion DM scenario.
		The edges of the polygon correspond to the minimal and maximal amplitudes which can be obtained for $\xi=100$ (solid lines) and $\xi=10$ (dashed lines), considering the case when the relaxion starts to oscillate immediately after barrier reappearance.
	}
	\label{fig:peak}
\end{figure}

The range of peak frequencies and amplitudes that can be attained for coherently oscillating relaxion DM is displayed in \cref{fig:peak}, considering the case when the relaxion starts oscillating at the time the wiggles reappear. 
The polygons are obtained from the allowed range for the cutoff $\Lambda$, for $\Tra$ between approximately $\SI{240}{\MeV}$ and \vEW; cf.~\cref{fig:Tra}. 
The solid and dashed lines assume $\xi=100$ and $\xi=10$, respectively.
Peak positions inside the polygons can be realized.
The sensitivities of \muAres, \experiment{LISA} and PTAs are again indicated as shaded regions.
Note that the shading and solid lines correspond to the sensitivity for $\xi=100$.
For $\xi=10$, the detection reach is degraded to the correspondingly coloured dashed lines.

For a large part of the peak frequencies and amplitudes that can be realized with relaxion DM, an observable signal is obtained, although mostly requiring futuristic space-based interferometers such as \muAres for observation. 
For low values of $\xi \sim 10$, a detection with \experiment{SKA} is possible. 
The present-day sensitivity of \experiment{NANOGrav} and expected reach of \experiment{LISA} are, however, insufficient for a detection.
While \experiment{NANOGrav} is able to exclude parts of the parameter space if the DM assumption is relaxed (cf.\ benchmark~A in \cref{fig:spectra}), the sensitivity of \experiment{LISA} will remain insufficient even in this more general case. 

We also show the GW sensitivity for relaxion DM as coloured regions in \cref{fig:DMGW}, using the same colouring scheme as above.
The (light) coloured regions enclosed by the solid green and blue lines here indicate the relaxion masses and mixing angles for which, at least in a subrange of the reappearance temperatures in accordance with the constraints, an observable signal in \muAres and \experiment{SKA} can be obtained.
While \muAres covers the full low-mass island as well as the range $\mrel \lesssim \SI{0.1}{\eV}$ in the high-mass island, the sensitivity of \experiment{SKA} is limited to slightly lower DM masses.
Note that, as the GW spectrum strongly depends on the reappearance temperature, a nonobservation in these experiments would in most cases not rule out the coloured parameter space, as a detection can be evaded by adjusting the reappearance temperature to shift the signal outside the experiment's reach.
In the dark-green coloured region bounded by the dotted lines, however, the generated stochastic GW background is observable in \muAres for the full range of allowed reappearance temperatures, guaranteeing a detectable signal in this region.
The temperature dependence of the relaxion constraints and GW limits can be seen explicitly in the animated version of the figure  that can be found in the ancillary material.

Last but not least, let us now dismiss the assumption that the relaxion constitutes DM and return to \cref{fig:RelaxionGW}, where we again indicate the parameter regions in which the GW background can be observed in \muAres~(green), \experiment{NANOGrav}~(purple), or \experiment{SKA}~(blue/turquoise) by the respective colouring.
The coloured regions respect the $\frel < \Mpl$ constraint.
If we allow for super-Planckian decay constant, the regions extend to the dotted lines.

Regarding the GW sensitivity, the reader needs to be aware that the figure shows the projection of a four-dimensional plot, as \Tra and \frel (or $\Lambda$) are not fixed.
While red and orange shading marks the values of \mrel and \srel for which it is not possible to evade the respective constraints by adjusting the remaining parameters (i.e.\ these coloured regions are definitely excluded), the GW contours (blue, turquoise, purple and green) correspond to the maximal reach of the respective experiments.
They are obtained by taking the maximal SNR that can be achieved in each experiment for the given values of \mrel and \srel.
Given the experimental sensitivities we assume here, a detection via GWs can be evaded in all of the viable parameter space.
In particular, the purple colouring and lines do not indicate that the corresponding parameter points are excluded by \experiment{NANOGrav} data but that \experiment{NANOGrav} constrains the reappearance temperature (and the decay constant) in this region.
The same also applies to the projections for \experiment{SKA} and \muAres.
Furthermore, although the sensitivities overlap in the plot, \muAres and PTAs operate in different frequency regimes and are therefore typically sensitive to different ranges of \Tra.
Nonetheless, we can conclude that current and future GW experiments can potentially detect the stochastic GWs from relaxion-stopping via dark photon emission, and thereby constrain the parameter space.
We again enclose an animated version of the figure in the ancillary material available online, where the dependence on \Tra (but not on $\Lambda$)  is shown.

\begin{figure}
	\centering
	\includegraphics[width=\columnwidth]{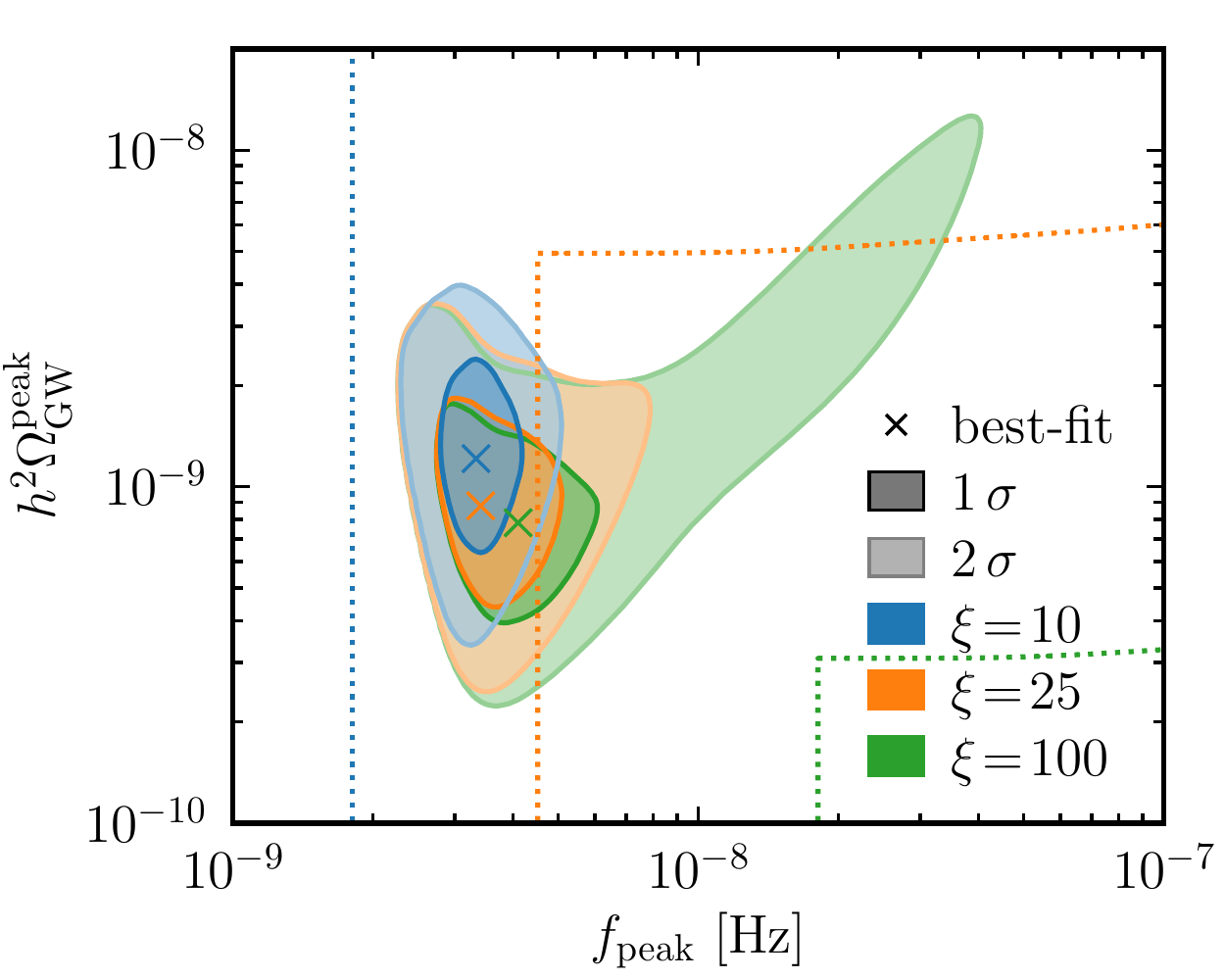}
	\caption{%
		Best-fit point~(cross) to the \experiment{NANOGrav} 12.5~year data as well as $1\sigma$~(dark) and $2\sigma$~(light) regions, fixing $\xi$ to 10~(blue), 25~(orange), and 100~(green).
		The dotted lines enclose the region that can be obtained without being in tension with $\Delta\Neff$ or BBN at the respective value of $\xi$.
	}
	\label{fig:NGfit}
\end{figure}

In addition to the current and prospective exclusion range, we also work out the parameter range in which our model can account for the potential GW signal recently observed in \experiment{NANOGrav}~\cite{Arzoumanian:2020vkk}.
Fitting our spectrum to the data using the same procedure as in Ref.~\cite{Ratzinger:2020koh}, where we keep $\xi$ fixed and only fit the peak frequency and amplitude, we obtain the best-fit points and the corresponding $1\sigma$ and $2\sigma$ contours shown in \cref{fig:NGfit}. 
We further indicate the minimal peak frequency dictated by the lower bound on the reappearance temperature, $\Tra \gtrsim T_\text{BBN}$, as well as the maximal peak amplitude consistent with the constraint on $\Delta\Neff$ by dotted lines.
While these bounds exclude an explanation of the observed stochastic process in terms of our model for large values of $\xi \sim 100$~(green), we can reach into the $1\sigma$ and $2\sigma$ regions for intermediate $\xi\sim 25$~(orange), and for $\xi\sim 10$~(blue), we can account for the \experiment{NANOGrav} best-fit point.

In the $\xi=10$ (lower) panel of \cref{fig:RelaxionGW}, we indicate the values of \mrel and \srel for which we can attain the best-fit point ($f_\text{peak} = \SI{3.3}{\nano\Hz}$, $h^2\Omega_\text{GW}^\text{peak} = \num{1.2e-9}$) by the grey shaded region.
Note that, as we fix $\xi$ and $f_\text{peak}$, this also fixes the reappearance temperature to $\Tra \sim \SI{20}{\MeV}$, while the peak amplitude then fixes \frel as a function of \mrel. 
Hence, in the grey shaded region, $\Lambda$ can be adjusted within the constraints to obtain the respective value of the mixing angle.
The best-fit spectrum for $\xi=10$ is also depicted in orange in \cref{fig:spectra}.

\section{Conclusion}
\label{sec:conclusion}

In this work, we have considered the possibility of probing the relaxion, which was proposed to ameliorate the Higgs hierarchy problem, via gravitational waves. 
A coupling to dark photons tames the relaxion excursion after reheating and thus in turn opens up a large fraction of the parameter space which was excluded in the minimal scenario without dark photons.
Furthermore, dark photon production after reheating can act as a source for the generation of a stochastic gravitational wave background.
The gravitational waves are sourced by the anisotropies that are induced by the tachyonic production of dark photons 
between the electroweak phase transition and big bang nucleosynthesis.
Hence, instead of the inflationary dynamics responsible for solving the hierarchy problem, we are probing here the late-time dynamics of the relaxion.

We have shown that this stochastic gravitational wave background can be probed by various current (\experiment{NANOGrav}) and future (\experiment{SKA} and \muAres) gravitational wave detectors. 
In addition, we also highlight the parameter range in which our gravitational wave signal can account for the common-spectrum process observed in the most recent \experiment{NANOGrav} data~\cite{Arzoumanian:2020vkk}. 
Alongside the existing theoretical constraints, we have presented the relaxion parameter space which can be detected or excluded by the gravitational wave observatories in \cref{fig:RelaxionGW}. 

We find that the spectral shape of the gravitational wave signal in our model falls as the fourth power of the frequency above the peak, unlike a steeper falling gravitational wave signals generated by other axionlike field dynamics~\cite{Machado:2018nqk,Machado:2019xuc,Ratzinger:2020oct,Co:2021rhi}, whereas it behaves like $f^3$ in the infrared, as expected based on causality arguments.
An observation of the spectrum in the range around the peak should allow for a determination of the reappearance temperature, while the amplitude can be used to determine the product of the relaxion mass and decay constant.

Furthermore, we have shown that the relaxion can constitute dark matter in the present Universe in the mass range of $\SI{e-11}{\eV}\lesssim \mrel\lesssim \SI{e-6}{\eV}$ and $\SI{e-2}{\eV} \lesssim \mrel \lesssim \SI{1}{\eV}$. 
While this scenario cannot be constrained with current \experiment{NANOGrav} data, most of the dark matter parameter space will be accessible by \experiment{SKA} and/or \muAres in the future. 
 Hence, with the advent of gravitational wave astronomy, we are now facing promising prospects for probing the relaxation of the electroweak scale via the stochastic gravitation wave background generated when stopping the relaxion, independently of whether the relaxion constitutes dark matter or not.

\begin{acknowledgments}
	We are grateful to Hyungjin Kim for collaboration at the early stages of this paper and participating in numerous vital discussions that seeded this work.
	We thank Edoardo Vitagliano for pointing out the updated fifth force bounds.
	The work of A.B.\ is supported by the Azrieli Foundation.  
	The work of E.M.\ is supported by the Minerva Foundation.
	The work of G.P.\ is supported by grants from BSF-NSF, Friedrich Wilhelm Bessel research award, GIF, ISF, Minerva, Yeda-Sela-SABRA-WRC, and the Segre Research Award. 
	The work of W.R.\ and P.S.\ is supported by the Deutsche Forschungsgemeinschaft (DFG), Project No.\ 438947057 and by the Cluster of Excellence “Precision Physics, Fundamental Interactions, and Structure of Matter” (PRISMA+ EXC 2118/1) funded by the German Research Foundation (DFG) within the German Excellence Strategy (Project No.\ 39083149).
\end{acknowledgments}

\appendix

\section{Minimal relaxion scenario}
\label{app:minimalScenario}

Here, we discuss the minimal relaxion scenario where a coupling to dark photons is absent. 
As discussed in \cref{sec:relaxion}, due to the restoration of the EW symmetry, the relaxion starts rolling again if the reheating temperature is higher than the EW scale, $\vEW$. 
To trap the relaxion in the minimum where it stopped during inflation, we need to require that it is displaced by less than $\Delta\theta \lesssim 2 \delta$, where $2 \delta$ is the separation between the minimum and the maximum~\cite{Banerjee:2020kww}. 
In the absence of a dark photon coupling, the Hubble friction is sufficient to trap the relaxion if $\mrel\lesssim \sqrt{8}\,\delta \Hra$, where \Hra is the Hubble scale at the time of barrier reappearance~\cite{Banerjee:2018xmn}.
In combination with the constraints discussed in \cref{eq:constraintslambdacosmo,eq:constraintsf,eq:constraintslTra}, this trapping condition limits the mixing angle as,
\begin{align}
\sin\theta_{h\phi}&\gtrsim \frac{\mrel^{3/2}\Lambda_{\rm min}}{m_h^2 (2\,\Hra^2)^{1/4}}\,\,\,&({\rm from}\,\, \Lambda\gtrsim\Lambda_{\rm min})\,,
\label{eq:minimalsthnodp}\\
\sin\theta_{h\phi}&\lesssim \frac{4 \,\Mpl\,\mrel\Hra}{m_h^2\, \vEW}\,\,\,&({\rm from}\,\, \frel\lesssim\Mpl)\,,
\label{eq:maximalsthnodp1}\\
\sin\theta_{h\phi}&\lesssim \frac{2^\frac{5}{4}\,\vEW \sqrt{\mrel\Hra}}{m_h^2 }\,\,\,&({\rm from}\,\, \Lbr\lesssim\vEW)\,,\label{eq:maximalsthnodp2}
\end{align}
within which the relaxion can be trapped by the Hubble friction. 
Moreover, the above equations restrict the relaxion mass for which the minimal scenario can be realized to $\mrel\lesssim \SI{2e-5}{\eV}$.

In addition to these constraints, for the relaxion not to overclose the Universe we need to require  $\Omega_{\phi}\lesssim\Omega_\text{DM}$.
This further constrains the range of the relaxion mass, and thus the minimal relaxion scenario can only be realized for 
\bea
\mrel\lesssim \SI{5e-8}{\eV}\,.
\eea

\begin{figure*}
	\centering
	\includegraphics[width=2\columnwidth]{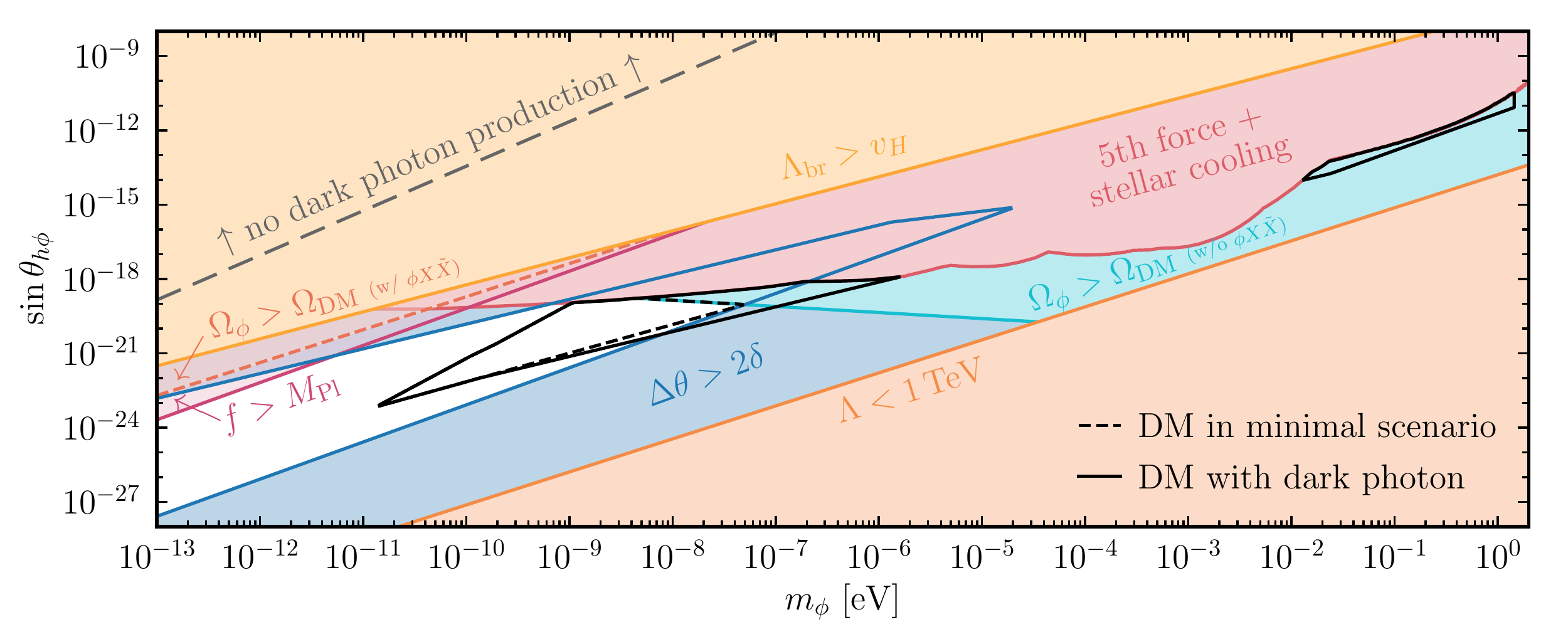}
	\caption{%
		Available parameter space in the minimal relaxion scenario when the reheating temperature is higher than the EW scale. 
		The red and orange shaded regions are excluded by the indicated constraints or combinations thereof.
		The blue shaded region is further excluded if the relaxion is not coupled to the dark photon source term, leading to a field excursion of more than $2\delta$. 
		The dark photon production cannot occur in the region above the grey dashed line. 
		Relaxion DM can be realized within the solid and dashed black contours, respectively, depending on whether a coupling to dark photons is assumed or not. 
		In the turquoise parameter region, the  condition $\Delta\theta \lesssim 2 \delta$ further leads to an overproduction of relaxion DM in the minimal scenario.
	}
	\label{fig:RelaxionParameterSpace}
\end{figure*}

\Cref{fig:RelaxionParameterSpace} shows the available parameter space in the relaxion mass \mrel vs.\ mixing angle \srel plane.
We assume here the minimal model without coupling to dark photons and do not require the relaxion to account for DM. 
Again, the red and orange shaded regions are excluded by the indicated constraints.
The condition that the relaxion remains trapped in the same minimum in which it settled before reheating, $\Delta\theta \lesssim 2 \delta$, then additionally excludes the region shaded in blue, with only the white region remaining as viable parameter space.
In the turquoise shaded region, fulfilling $\Delta\theta \lesssim 2 \delta$ without dark photons further leads to an overclosure of the Universe.
The region in which the relaxion can constitute DM is enclosed by the dashed black contour.
If the relaxion is coupled to a dark photon, we further open up the blue and turquoise shaded regions of the parameter space.
With a dark photon, relaxion DM can be realized within the solid black contour (same as the black framed regions of \cref{fig:DMGW}). 
Note that the dark photon scenario can only be applied in the region below the grey dashed line.
Above the line, the condition $\Hpp > \Hra$ required for dark photon production is not satisfied (cf.\ the discussion in \cref{sec:constraints}).

\section{Dark photon spectrum}
\label{app:DPspectrum}

As discussed in \cref{sec:darkphoton}, since $\theta' > 0$ in our case, positive-helicity dark photon modes with $k < \ke = \rX \theta'$ are tachyonic and experience exponential growth, whereas modes outside the tachyonic band oscillate.
In particular, the amplitude of the oscillating modes does not change in time.
At any time, the peak of the dark photon spectrum is given by the mode that experiences maximal growth with $k = \km(\tau) = |\rX \theta'|/2$.
We therefore take the ansatz
\begin{equation}
	X_+(k,\tau) = \begin{cases}
		\normk \cos\left(k\tau - \xi\right)
		& \text{for}\ \km < k < \kpp ,
		\\
		0 & \text{otherwise},
	\end{cases}
\end{equation}
where we neglect the negative-helicity modes as well as all modes that did not experience maximal growth yet, since these are exponentially suppressed.
We treat modes with $\ke < k < \km$ like the oscillating modes in order to include the peak of the spectrum in our estimate.

Neglecting the energy produced before particle production, the dark photon energy density is given by
\begin{equation}
	\rho_X = \int\limits_{\km}^{\kpp}\! \frac{d k\, k^2}{4 \pi^2 a^4} \left(
	|X'|^2 
	+ k^2 |X|^2 
	\right) 
	= \!\int\limits_{\km}^{\kpp}\!\! \frac{d k}{k}\, \frac{|\normk|^2 k^5}{4 \pi^2 a^4} ,
\end{equation}
where $\kpp=\ke(\tapp)$ is the mode that exits the tachyonic band at particle production.
On the other hand, we can determine the energy density  spectrum of the dark photons from the relaxion energy density as
\begin{equation}\begin{aligned}
		\rho_X 
		&= -\! \int\limits_{\tapp}^{\tau}\!d\eta\,\frac{a^4(\eta)}{a^4(\tau)} \frac{\partial V}{\partial \theta}\,\frac{\partial\theta}{\partial \eta} 
		\approx \int\limits_{\km}^{\kpp}\!\frac{d k}{k}\, \frac{\km^4}{k^4} \frac{\xi \Lbr^4}{\rX}\,,	
\end{aligned}\end{equation}
where we assumed that at each time energy is dominantly transferred into the maximally growing mode, which goes like $\km \propto 1/\tau$.
Hence, we can rewrite $\normk = \normX k^{-9/2}$ with
\begin{equation}
	\normX = \frac{\pi}{2}\, \Lbr^2\,\sqrt{\frac{\xi}{\rX}} \,\ara^2 \kra^2 \,,
\end{equation}
where \kra is the mode that exits the tachyonic band at reappearance, $\kra = \xi/\tara$.
The dark photon spectrum therefore becomes
\begin{equation}
	\label{eq:Xspectrum}
	X_+(k,\tau) = \begin{cases}
		\normX k^{-\frac{9}{2}} \cos\left(k\tau - \xi\right)
		& \text{for}\ \km < k < \kpp ,
		\\
		0 & \text{otherwise}.
	\end{cases}
\end{equation}

\begin{figure}
	\centering
	\includegraphics[width=\columnwidth]{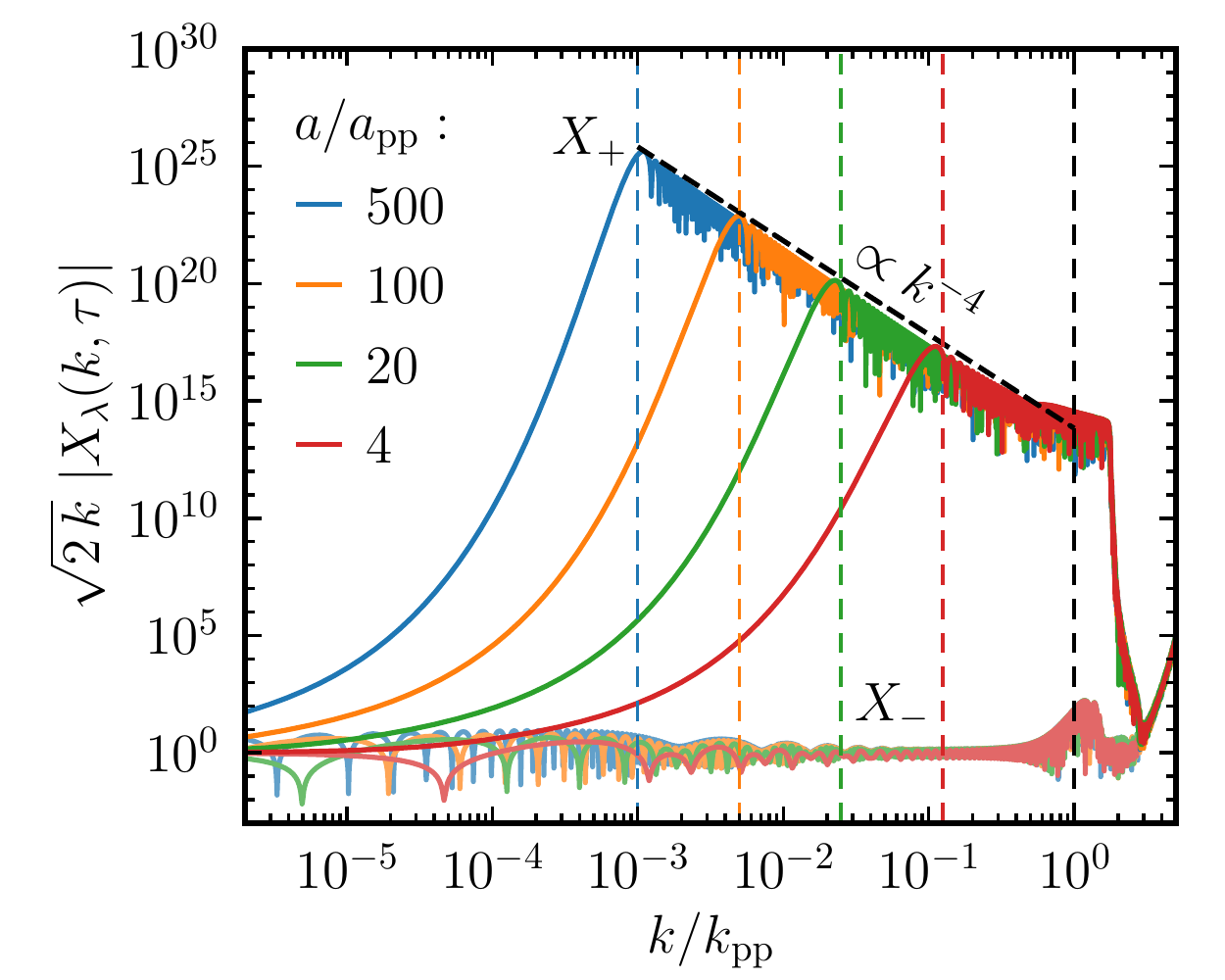}
	\caption{%
		Simulated (coloured lines) and expected (dashed black line) amplitude of the dark photon modes at different times.
		The deep (light) coloured lines correspond to the positive (negative) helicity.
		At each time, the expected peak momentum is indicated by the vertical dashed line in the corresponding colour, whereas the black-dashed vertical line indicates the upper bound $k < \kpp$ on the tachyonic dark photon momentum.
	}
	\label{fig:DPspectrum}
\end{figure}

To corroborate our estimation, we have simulated the dark photon and relaxion evolution after reheating, solving the equations of motion~\labelcref{eq:EOMrelaxion,eq:EOMdarkphoton} numerically, using \num{5000} logarithmically spaced, discretized momenta.
In \cref{fig:DPspectrum}, we present the result of the numeric simulation for the dark photon modes as a function of the momentum $k$.
The simulation assumes $\mrel =   \SI{1}{\meV}$, $\frel = \SI{2.35e13}{\GeV}$, $\Lambda = \SI{100}{\TeV}$, and $\xi = 77$.
The value of $\rX$ has been determined numerically from \cref{eq:XXtilde}~\cite{Banerjee:2018xmn}.
We show the spectrum at $a/\app = 4$~(red), $20$~(green), $100$~(orange), and $500$~(blue), where the full and light coloured lines correspond to the positive and negative helicity, respectively.
As expected, the positive helicity modes dominate over the negative helicity by far.
Furthermore, the amplitude for positive helicities indeed follows a $k^{-9/2}$ law [cf.~\cref{eq:Xspectrum}, i.e.~$\sqrt{2 k}|X_+|$ goes as $k^{-4}$] between the momentum \kpp that exits the tachyonic band at particle production and the peak momentum $\km(\tau)$ that experiences the largest growth rate at the respective time.
The peak momenta and $\kpp$ are indicated by the coloured and black, dashed vertical lines, respectively.

\section{Calculation of the gravitational wave spectrum}
\label{app:GWcalculation}

The energy density of a stochastic GW background is given by
\begin{equation}
\begin{aligned}
	\rhoGW(\vect{x},\tau) &= \frac{\Mpl^2}{4\,a^2(\tau)} \left<\hat{h}'_{ab}(\vect{x},\tau) \hat{h}^{\prime ab}(\vect{x},t)\right>\\
	&= \frac{\Mpl^2}{8\pi^2 a^2(\tau)} \int\!d k\,k^2 \Ph(k,\tau) \,,
\end{aligned}
\end{equation}
where \Ph is the power spectral density of $h'_{ab}$ defined by $\langle \hat{h}^{\prime ab}(\vect{k},\tau) \hat{h}^{\prime\ast}_{ab}(\vect{k}',\tau) \rangle = (2\pi)^3 \delta(\vect{k}-\vect{k}') \Ph(k,\tau)$.
We hence obtain the GW spectrum
\begin{equation}
	\label{eq:rhoGW}
	\frac{d \rhoGW}{d \log k}(k,\tau) = \frac{\Mpl^2 k^3}{8 \pi^2 a^2(\tau)} \Ph(k,\tau) \,.
\end{equation}
Using the solution \cref{eq:hij}, we obtain
\begin{multline}
	\label{eq:psd}
	\Ph(k,\tau) = \frac{2}{\Mpl^4 a^2(\tau)} \int\limits_{\tapp}^{\tau}\!\!d\tau' \!\!\int\limits_{\tapp}^{\tau}\!\!d\tau''\, \cos[k(\tau'-\tau'')] \,
	\times\\
	a(\tau')\, a(\tau'')\, \Pi^2(k,\tau',\tau'')
\end{multline}
with the unequal time correlator again defined as
$\langle0| \hat{\Pi}^{ab}(\vect{k},\tau) \hat{\Pi}^{\ast}_{ab}(\vect{k}',\tau')|0 \rangle = (2\pi)^3 \delta(\vect{k}-\vect{k}') \Pi^2(k,\tau,\tau')$,
where we averaged over one period in $\tau$ to get rid of the $\cos[k(2\tau-\tau'-\tau'')]$ part.

Inserting \cref{eq:Piab,eq:WKB} into \cref{eq:psd,eq:rhoGW} and rewriting the dark electric and magnetic fields in terms of the dark photon modes,
\begin{subequations}
	\label{eq:EandB}
	\begin{align}
		\hat{\vect{E}}(\vect{k},\tau) &= \sum\limits_{\lambda} X'_\lambda(k,\tau)\,\vect{\varepsilon}_{\lambda}(\vect{k})\,\hat{a}_\lambda(\vect{k}) +\hc,\\
		\hat{\vect{B}}(\vect{k},\tau) &= \sum\limits_{\lambda} \lambda\,k\,X_\lambda(k,\tau)\,\vect{\varepsilon}_{\lambda}(\vect{k})\,\hat{a}_\lambda(\vect{k}) +\hc,
	\end{align}
\end{subequations}
we obtain the spectrum at the time of reappearance as
\begin{multline}
	\label{eq:rhoGWIcs}
	\frac{d \rhoGW}{d \log k} = 
	\frac{k^3}{2\pi^2 \Mpl^2 \ara^4} 
	\int\!\frac{d^3 q}{(2\pi)^3}\, 
	|\Sigma_{ab}(\vect{k},\vect{q})|^2 \times \\
	\left(
	|\Ics{c}(\vect{k},\vect{q})|^2 + |\Ics{s}(\vect{k},\vect{q})|^2
	\right) \,,
\end{multline}
where we used $[\hat{a}_\lambda(\vect{k}),\hat{a}_{\lambda'}^\dagger(\vect{k}')] = (2\pi)^3 \delta_{\lambda\lambda'} \delta(\vect{k}-\vect{k}')$ and the invariance of the integrand upon interchanging $\vect{q}\to\vect{k}-\vect{q}$.
We only considered here the positive helicity modes and defined 
$\Sigma_{ab}(\vect{k},\vect{q}) = \Lambda_{ab}^{ij}(\vect{k}) \varepsilon_i^+(\vect{q}) \varepsilon_j^+(\vect{k}-\vect{q}) $
as well as
\begin{equation}
	\Ics{c/s}(\vect{k},\vect{q}) = - \int\limits_{\tapp}^{\tara} \!\frac{d\tau}{a(\tau)}\,  
	\left\{ \begin{array}{l}
		\cos(k \tau) \\
		\sin(k\tau) \rule{0pt}{12pt}
	\end{array} \right\}
	\chi(q,l,\tau)
\end{equation}
with $\chi(\vect{k},\vect{q},\tau) = X_+'(q,\tau) X_+'(l,\tau) + q X_+(q,\tau) \, l X_+(l,\tau)$
and $l = |\vect{k}-\vect{q}|$.
Using the dark photon spectrum \cref{eq:Xspectrum} we obtain
\begin{equation}
	\chi(q,l,\tau) = \frac{\normX^2 \cos[(q-l)\tau] }{(q l)^\frac{7}{2}} 
\end{equation}
for $\min(q,l) > \km(\tau)$.

Expressing the momenta in terms of $\kra$, we can rewrite the energy density as
\begin{widetext}
\begin{align}
	\label{eq:GWspectrumExact}
	\frac{d \rhoGW}{d \log k} &=  \frac{\Lbr^8 \xi^2}{\rX^2 \Hra^2 \Mpl^2}\ x^3 \int\limits_\frac{1}{2}^\infty\! d r \int\limits_{-1}^1 \!d\cos\theta \, \frac{|\Sigma_{ab}(x,r,\cos\theta)|^2}{512\, r^5 s^7} \left( |\Icst{c}(x,r,s, \xi)|^2 + |\Icst{s}(x,r,s, \xi)|^2\right) \ \Theta\left(s-1/2\right)\,,
\end{align}
\end{widetext}
where the remaining integrals only depend on $x = k/\kra$ and $\xi$. 
We defined here $r=q/\kra$ and $s=l/\kra$. 
The polarization part is given by (see, e.g., Ref.~\cite{Machado:2018nqk} for further details)
\begin{align}
	\label{eq:GWpolarization}
	|\Sigma_{ab}|^2 = \sum\limits_{\lambda=\pm} \left[\frac{1 + \lambda \cos\theta}{2}\right]^2  \left[\frac{1 + \lambda \frac{x - r  \cos\theta}{s}}{2}\right]^2 \!\!\!,	
\end{align}
where $\lambda$ now denotes the GW helicity, 
and the time integrals evaluate to
\begin{align}
	\label{eq:Ics}
	\Icst{c/s}\! =\! \begin{cases} 
		\!\CosInt\!\left[(x+r-s)\xi\right] - \CosInt\!\left[\frac{(x+r-s)\xi}{2 \,\min(r,s)}\right] \!+ ( r \leftrightarrow s) \\
		\SinInt\!\left[(x+r-s)\xi\right] - \,\SinInt\!\left[\frac{(x+r-s)\xi}{2 \,\min(r,s)}\right] \!+ ( r \leftrightarrow s) \rule{0pt}{14pt}
	\end{cases}
	\hspace*{-15pt}
\end{align}
where $\CosInt$ and $\SinInt$ are the cosine and sine integral function, 
\begin{align}
	\CosInt(z) = \int\limits_z^\infty \! d t\, \frac{\cos t}{t} \,,\qquad
	\SinInt(z) = \int\limits_0^z \! d t\, \frac{\sin t}{t}\,.
\end{align}

The GW spectrum can now be obtained by evaluating \cref{eq:GWspectrumExact} numerically. 
In this paper, we, however, use an analytic approximation to the spectrum based on the amplitude at the peak and the asymptotic behaviour in the UV and IR.

Since the GW momentum is given by the sum of two dark photon momenta, $\vect{k} = \vect{q}+\vect{l}$, and since the time integrals are dominated by the late-time behaviour, the peak of the GW spectrum will roughly be given by twice the peak momentum of the dark photon spectrum at reappearance, i.e.\ we take $k_\text{peak} = \kra$. 
As the arguments of the cosine and sine integrals in \cref{eq:Ics}  are proportional to $\xi \sim \order{10-100}$, we can expand for large $\xi$. 
Assuming that the cosine and sine terms remaining in the expansion oscillate quickly and therefore average to zero, the corresponding amplitude for $x=1$ then evaluates to%
\footnote{The analytic integration actually yields $\frac{\num{278880340} + \num{44998983} \log 3}{\num{1254113280}}$ $\approx 0.262$ instead of $1/4$.}
\begin{align}
	\label{eq:GWspectrumPeak}
	\frac{d \rhoGW^\text{peak}}{d \log k} = \left[\frac{1}{4\,\xi^2} + \order{\xi^{-4}} \right] \frac{\xi^2}{\rX^2} \frac{\Lbr^8}{\Hra^2 \Mpl^2} \,.
\end{align}
For $k \gg \kra$, we have $x+r-s = r (1+\cos\theta) + \order{x^{-1}}$ and $x-r+s = 2 x + \order{1}$.
The $(r \leftrightarrow s)$ part of \cref{eq:Ics} is hence suppressed by $1/x$, and only the first two terms contribute. 
Again expanding for large $\xi$, we obtain that the spectrum behaves as
\begin{align}
	\label{eq:GWspectrumUV}
	\frac{d \rhoGW^\text{UV}}{d \log k} =  \left[\frac{5}{192\,\xi^2} + \order{\xi^{-4}} \right] \frac{\xi^2}{\rX^2} \frac{\Lbr^8}{\Hra^2 \Mpl^2} \frac{\kra^4}{k^4}
\end{align}
in the UV.
For $k \ll \kra$, on the other hand, the arguments of the cosine and sine integrals become proportional to $x \pm r \mp s = x (1 \pm \cos\theta) + \order{x^2}$, so we can expand the integrals for low arguments, yielding the IR behaviour
\begin{align}
	\label{eq:GWspectrumIR}
	\frac{d \rhoGW^\text{IR}}{d \log k} = \frac{64}{\num{19965}} \frac{\xi^2}{\rX^2} \frac{\Lbr^8}{\Hra^2 \Mpl^2} \frac{k^3}{\kra^3}\,.
\end{align}
As can be seen from \cref{eq:GWpolarization}, at low momentum, $x \ll1$, where we have $s \sim r$, both GW helicities contribute equally, whereas at high momentum, where $x \gg 1$ and hence $s \sim x$, the negative helicity is suppressed by a factor of $1/k^2$ compared to the positive one.

Our approximation to the full spectrum is now obtained by combining these results inversely. 
Neglecting GW production after reappearance, the energy density will subsequently simply redshift as $\rhoGW\sim a^{-4}$, so the spectrum today as a function of the comoving wave number $k$ becomes
\begin{widetext}
\begin{align}
	\label{eq:GWspectrumApproximation}
	\OGW(k)  =
	 \frac{1}{\rhoc} \left(\frac{\ara}{a_0}\right)^4 \!\frac{1}{\Mpl^2 \Hra^2} \frac{\Lbr^8}{4\,\rX^2} \left\{
	 	1 + 
	 	\frac{\num{19965}}{256\,\xi^2} \left[ \left( \frac{\kra^3}{k^3} - 1\right) +   3 \left( \frac{k}{\kra} - 1\right)  \right]
	 	+ \frac{48}{5} \left( \frac{k}{\kra} - 1\right)^4
	 \right\}^{-1} \hspace{-6pt},
\end{align}
\end{widetext}
where the  first term in the curly bracket reproduces our estimate for the peak amplitude; the last term and the first part of the second term reproduce the spectrum in the IR and UV, respectively; and the remaining term in the square bracket renders $k = \kra$ an extremum of the spectrum.
Further noting that frequency $f$ and comoving wave number $k$ are related via
$k d \tau = \frac{k}{a} d t = 2 \pi \freq d t$, 
we end up with the present-day spectrum in \cref{eq:GWspectrum,eq:fPeak,eq:OmegaPeak,eq:SpectralShape}.

\begin{figure}
	\includegraphics[width=\columnwidth]{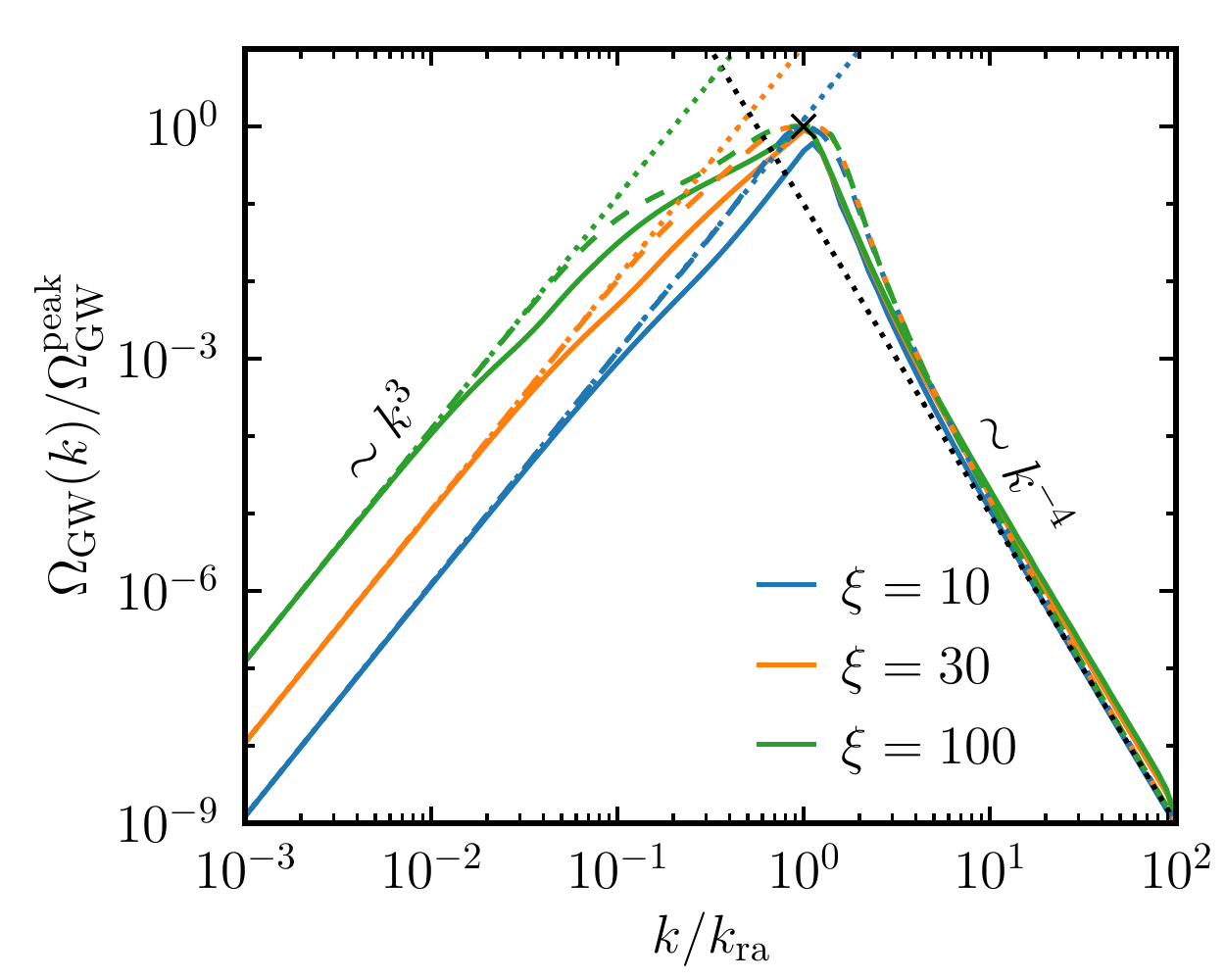}
	\caption{%
		Spectral shape of the stochastic GW background for $\xi=10$ (blue), $30$ (orange), and $100$ (green).
		The solid lines correspond to the numeric calculation, whereas the dashed ones are the analytic approximation.
		The dotted lines indicate the IR and UV behaviour, and the black cross marks our estimate  of the peak amplitude.
	}
	\label{fig:spectralShape}
\end{figure}

In \cref{fig:spectralShape}, we compare the numerical result of the spectral shape based on \cref{eq:GWspectrumExact} (solid lines) to the analytic approximation in \cref{eq:GWspectrumApproximation} (dashed lines) for different values of $\xi$. 
In addition, the UV and IR behaviour \cref{eq:GWspectrumUV,eq:GWspectrumIR} as well as the peak amplitude \cref{eq:GWspectrumPeak} are indicated by dotted lines and a cross, respectively.
Our approximation agrees well with the numeric result. 

\begin{figure}
	\includegraphics[width=\columnwidth]{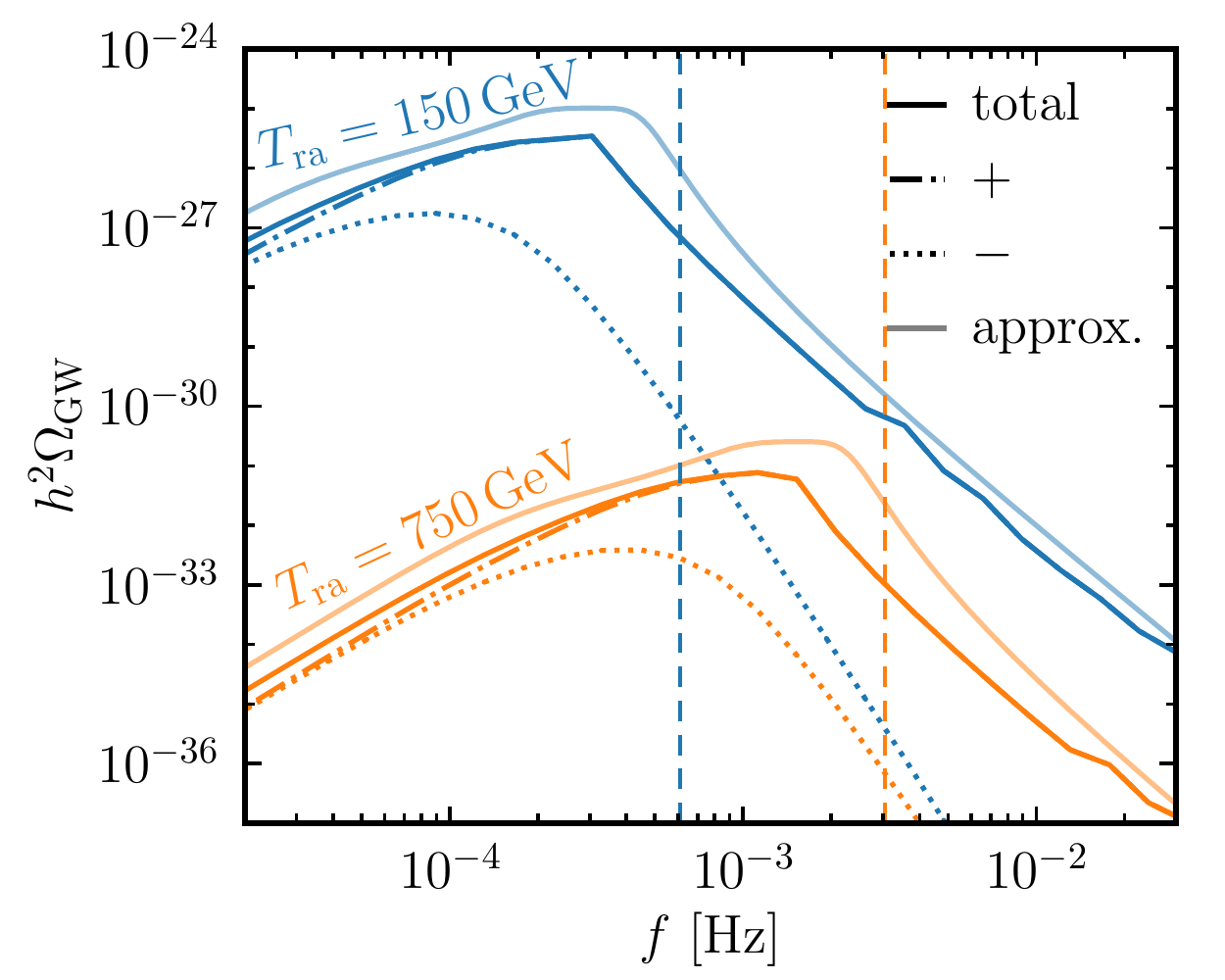}
	\caption{%
		Simulation (full colours) and analytic approximation (light colours) of the GW spectrum for a reappearance temperature of $\Tra = \SI{150}{\GeV}$~(blue) and \SI{750}{\GeV}~(orange).
		The vertical dashed lines indicate the expected peak frequency, whereas the dot-dashed and dotted curves correspond to the positive and negative helicity contributions to the simulated spectrum.
	}
	\label{fig:simulation}
\end{figure}

Finally, \cref{fig:simulation} shows the GW spectrum from our numeric simulation (cf.~\cref{fig:DPspectrum}), assuming a reappearance temperature of $\Tra = \SI{150}{\GeV}$~(blue) and \SI{750}{\GeV}~(orange), corresponding to $a/\ara = 500$ and $a/\ara=100$, respectively.
We see that the GW background is unpolarized below the peak, whereas above the peak the positive polarization~(dot-dashed) dominates over the negative one~(dotted).
Our estimates based on \cref{eq:GWspectrumApproximation}~(light lines) overestimate the simulated spectra by an \order{1-10} factor.

\bibliography{relaxionGW.bib}

\begin{thebibliography}{60}%
\makeatletter
\providecommand \@ifxundefined [1]{%
 \@ifx{#1\undefined}
}%
\providecommand \@ifnum [1]{%
 \ifnum #1\expandafter \@firstoftwo
 \else \expandafter \@secondoftwo
 \fi
}%
\providecommand \@ifx [1]{%
 \ifx #1\expandafter \@firstoftwo
 \else \expandafter \@secondoftwo
 \fi
}%
\providecommand \natexlab [1]{#1}%
\providecommand \enquote  [1]{``#1''}%
\providecommand \bibnamefont  [1]{#1}%
\providecommand \bibfnamefont [1]{#1}%
\providecommand \citenamefont [1]{#1}%
\providecommand \href@noop [0]{\@secondoftwo}%
\providecommand \href [0]{\begingroup \@sanitize@url \@href}%
\providecommand \@href[1]{\@@startlink{#1}\@@href}%
\providecommand \@@href[1]{\endgroup#1\@@endlink}%
\providecommand \@sanitize@url [0]{\catcode `\\12\catcode `\$12\catcode
  `\&12\catcode `\#12\catcode `\^12\catcode `\_12\catcode `\%12\relax}%
\providecommand \@@startlink[1]{}%
\providecommand \@@endlink[0]{}%
\providecommand \url  [0]{\begingroup\@sanitize@url \@url }%
\providecommand \@url [1]{\endgroup\@href {#1}{\urlprefix }}%
\providecommand \urlprefix  [0]{URL }%
\providecommand \Eprint [0]{\href }%
\providecommand \doibase [0]{https://doi.org/}%
\providecommand \selectlanguage [0]{\@gobble}%
\providecommand \bibinfo  [0]{\@secondoftwo}%
\providecommand \bibfield  [0]{\@secondoftwo}%
\providecommand \translation [1]{[#1]}%
\providecommand \BibitemOpen [0]{}%
\providecommand \bibitemStop [0]{}%
\providecommand \bibitemNoStop [0]{.\EOS\space}%
\providecommand \EOS [0]{\spacefactor3000\relax}%
\providecommand \BibitemShut  [1]{\csname bibitem#1\endcsname}%
\let\auto@bib@innerbib\@empty
\bibitem [{\citenamefont {Caprini}\ and\ \citenamefont
  {Figueroa}(2018)}]{Caprini:2018mtu}%
  \BibitemOpen
  \bibfield  {author} {\bibinfo {author} {\bibfnamefont {C.}~\bibnamefont
  {Caprini}}\ and\ \bibinfo {author} {\bibfnamefont {D.~G.}\ \bibnamefont
  {Figueroa}},\ }\bibfield  {title} {\bibinfo {title} {{Cosmological
  backgrounds of gravitational waves}},\ }\href
  {https://doi.org/10.1088/1361-6382/aac608} {\bibfield  {journal} {\bibinfo
  {journal} {Class. Quant. Grav.}\ }\textbf {\bibinfo {volume} {35}},\ \bibinfo
  {pages} {163001} (\bibinfo {year} {2018})},\ \Eprint
  {https://arxiv.org/abs/1801.04268} {arXiv:1801.04268 [astro-ph.CO]}
  \BibitemShut {NoStop}%
\bibitem [{\citenamefont {Caprini}\ \emph {et~al.}(2020)\citenamefont {Caprini}
  \emph {et~al.}}]{Caprini:2019egz}%
  \BibitemOpen
  \bibfield  {author} {\bibinfo {author} {\bibfnamefont {C.}~\bibnamefont
  {Caprini}} \emph {et~al.},\ }\bibfield  {title} {\bibinfo {title} {{Detecting
  gravitational waves from cosmological phase transitions with LISA: an
  update}},\ }\href {https://doi.org/10.1088/1475-7516/2020/03/024} {\bibfield
  {journal} {\bibinfo  {journal} {J. Cosmol. Astropart. Phys.}\ }\textbf
  {\bibinfo {volume} {2020}}\bibfield  {number} {\bibinfo  {number} { (03)},\
  \bibinfo {pages} {024}},\ }\Eprint {https://arxiv.org/abs/1910.13125}
  {arXiv:1910.13125 [astro-ph.CO]} \BibitemShut {NoStop}%
\bibitem [{\citenamefont {Graham}\ \emph {et~al.}(2015)\citenamefont {Graham},
  \citenamefont {Kaplan},\ and\ \citenamefont {Rajendran}}]{Graham:2015cka}%
  \BibitemOpen
  \bibfield  {author} {\bibinfo {author} {\bibfnamefont {P.~W.}\ \bibnamefont
  {Graham}}, \bibinfo {author} {\bibfnamefont {D.~E.}\ \bibnamefont {Kaplan}},\
  and\ \bibinfo {author} {\bibfnamefont {S.}~\bibnamefont {Rajendran}},\
  }\bibfield  {title} {\bibinfo {title} {{Cosmological Relaxation of the
  Electroweak Scale}},\ }\href {https://doi.org/10.1103/PhysRevLett.115.221801}
  {\bibfield  {journal} {\bibinfo  {journal} {Phys. Rev. Lett.}\ }\textbf
  {\bibinfo {volume} {115}},\ \bibinfo {pages} {221801} (\bibinfo {year}
  {2015})},\ \Eprint {https://arxiv.org/abs/1504.07551} {arXiv:1504.07551
  [hep-ph]} \BibitemShut {NoStop}%
\bibitem [{\citenamefont {Salopek}\ \emph {et~al.}(1989)\citenamefont
  {Salopek}, \citenamefont {Bond},\ and\ \citenamefont
  {Bardeen}}]{Salopek:1988qh}%
  \BibitemOpen
  \bibfield  {author} {\bibinfo {author} {\bibfnamefont {D.~S.}\ \bibnamefont
  {Salopek}}, \bibinfo {author} {\bibfnamefont {J.~R.}\ \bibnamefont {Bond}},\
  and\ \bibinfo {author} {\bibfnamefont {J.~M.}\ \bibnamefont {Bardeen}},\
  }\bibfield  {title} {\bibinfo {title} {{Designing density fluctuation spectra
  in inflation}},\ }\href {https://doi.org/10.1103/PhysRevD.40.1753} {\bibfield
   {journal} {\bibinfo  {journal} {Phys. Rev. D}\ }\textbf {\bibinfo {volume}
  {40}},\ \bibinfo {pages} {1753} (\bibinfo {year} {1989})}\BibitemShut
  {NoStop}%
\bibitem [{\citenamefont {B\"odeker}\ and\ \citenamefont
  {Buchm\"uller}(2020)}]{Bodeker:2020ghk}%
  \BibitemOpen
  \bibfield  {author} {\bibinfo {author} {\bibfnamefont {D.}~\bibnamefont
  {B\"odeker}}\ and\ \bibinfo {author} {\bibfnamefont {W.}~\bibnamefont
  {Buchm\"uller}},\ }\bibfield  {title} {\bibinfo {title} {{Baryogenesis from
  the weak scale to the grand unification scale}},\ }\href@noop {} {\
  (\bibinfo {year} {2020})},\ \Eprint {https://arxiv.org/abs/2009.07294}
  {arXiv:2009.07294 [hep-ph]} \BibitemShut {NoStop}%
\bibitem [{\citenamefont {Banerjee}\ \emph {et~al.}(2019)\citenamefont
  {Banerjee}, \citenamefont {Kim},\ and\ \citenamefont
  {Perez}}]{Banerjee:2018xmn}%
  \BibitemOpen
  \bibfield  {author} {\bibinfo {author} {\bibfnamefont {A.}~\bibnamefont
  {Banerjee}}, \bibinfo {author} {\bibfnamefont {H.}~\bibnamefont {Kim}},\ and\
  \bibinfo {author} {\bibfnamefont {G.}~\bibnamefont {Perez}},\ }\bibfield
  {title} {\bibinfo {title} {{Coherent relaxion dark matter}},\ }\href
  {https://doi.org/10.1103/PhysRevD.100.115026} {\bibfield  {journal} {\bibinfo
   {journal} {Phys. Rev. D}\ }\textbf {\bibinfo {volume} {100}},\ \bibinfo
  {pages} {115026} (\bibinfo {year} {2019})},\ \Eprint
  {https://arxiv.org/abs/1810.01889} {arXiv:1810.01889 [hep-ph]} \BibitemShut
  {NoStop}%
\bibitem [{\citenamefont {Banerjee}\ \emph {et~al.}(2020)\citenamefont
  {Banerjee}, \citenamefont {Kim}, \citenamefont {Matsedonskyi}, \citenamefont
  {Perez},\ and\ \citenamefont {Safronova}}]{Banerjee:2020kww}%
  \BibitemOpen
  \bibfield  {author} {\bibinfo {author} {\bibfnamefont {A.}~\bibnamefont
  {Banerjee}}, \bibinfo {author} {\bibfnamefont {H.}~\bibnamefont {Kim}},
  \bibinfo {author} {\bibfnamefont {O.}~\bibnamefont {Matsedonskyi}}, \bibinfo
  {author} {\bibfnamefont {G.}~\bibnamefont {Perez}},\ and\ \bibinfo {author}
  {\bibfnamefont {M.~S.}\ \bibnamefont {Safronova}},\ }\bibfield  {title}
  {\bibinfo {title} {{Probing the relaxed relaxion at the luminosity and
  precision frontiers}},\ }\href {https://doi.org/10.1007/JHEP07(2020)153}
  {\bibfield  {journal} {\bibinfo  {journal} {J. High Energy Phys.}\ }\textbf
  {\bibinfo {volume} {2020}}\bibfield  {number} {\bibinfo  {number} { (07)},\
  \bibinfo {pages} {153}},\ }\Eprint {https://arxiv.org/abs/2004.02899}
  {arXiv:2004.02899 [hep-ph]} \BibitemShut {NoStop}%
\bibitem [{\citenamefont {Choi}\ and\ \citenamefont
  {Im}(2016{\natexlab{a}})}]{Choi:2016luu}%
  \BibitemOpen
  \bibfield  {author} {\bibinfo {author} {\bibfnamefont {K.}~\bibnamefont
  {Choi}}\ and\ \bibinfo {author} {\bibfnamefont {S.~H.}\ \bibnamefont {Im}},\
  }\bibfield  {title} {\bibinfo {title} {{Constraints on relaxion windows}},\
  }\href {https://doi.org/10.1007/JHEP12(2016)093} {\bibfield  {journal}
  {\bibinfo  {journal} {J. High Energy Phys.}\ }\textbf {\bibinfo {volume}
  {2016}}\bibfield  {number} {\bibinfo  {number} { (12)},\ \bibinfo {pages}
  {093}},\ }\Eprint {https://arxiv.org/abs/1610.00680} {arXiv:1610.00680
  [hep-ph]} \BibitemShut {NoStop}%
\bibitem [{\citenamefont {Flacke}\ \emph {et~al.}(2017)\citenamefont {Flacke},
  \citenamefont {Frugiuele}, \citenamefont {Fuchs}, \citenamefont {Gupta},\
  and\ \citenamefont {Perez}}]{Flacke:2016szy}%
  \BibitemOpen
  \bibfield  {author} {\bibinfo {author} {\bibfnamefont {T.}~\bibnamefont
  {Flacke}}, \bibinfo {author} {\bibfnamefont {C.}~\bibnamefont {Frugiuele}},
  \bibinfo {author} {\bibfnamefont {E.}~\bibnamefont {Fuchs}}, \bibinfo
  {author} {\bibfnamefont {R.~S.}\ \bibnamefont {Gupta}},\ and\ \bibinfo
  {author} {\bibfnamefont {G.}~\bibnamefont {Perez}},\ }\bibfield  {title}
  {\bibinfo {title} {{Phenomenology of relaxion-Higgs mixing}},\ }\href
  {https://doi.org/10.1007/JHEP06(2017)050} {\bibfield  {journal} {\bibinfo
  {journal} {J. High Energy Phys.}\ }\textbf {\bibinfo {volume}
  {2017}}\bibfield  {number} {\bibinfo  {number} { (06)},\ \bibinfo {pages}
  {050}},\ }\Eprint {https://arxiv.org/abs/1610.02025} {arXiv:1610.02025
  [hep-ph]} \BibitemShut {NoStop}%
\bibitem [{\citenamefont {Choi}\ \emph {et~al.}(2017)\citenamefont {Choi},
  \citenamefont {Kim},\ and\ \citenamefont {Sekiguchi}}]{Choi:2016kke}%
  \BibitemOpen
  \bibfield  {author} {\bibinfo {author} {\bibfnamefont {K.}~\bibnamefont
  {Choi}}, \bibinfo {author} {\bibfnamefont {H.}~\bibnamefont {Kim}},\ and\
  \bibinfo {author} {\bibfnamefont {T.}~\bibnamefont {Sekiguchi}},\ }\bibfield
  {title} {\bibinfo {title} {{Dynamics of the cosmological relaxation after
  reheating}},\ }\href {https://doi.org/10.1103/PhysRevD.95.075008} {\bibfield
  {journal} {\bibinfo  {journal} {Phys. Rev. D}\ }\textbf {\bibinfo {volume}
  {95}},\ \bibinfo {pages} {075008} (\bibinfo {year} {2017})},\ \Eprint
  {https://arxiv.org/abs/1611.08569} {arXiv:1611.08569 [hep-ph]} \BibitemShut
  {NoStop}%
\bibitem [{\citenamefont {Hook}\ and\ \citenamefont
  {Marques-Tavares}(2016)}]{Hook:2016mqo}%
  \BibitemOpen
  \bibfield  {author} {\bibinfo {author} {\bibfnamefont {A.}~\bibnamefont
  {Hook}}\ and\ \bibinfo {author} {\bibfnamefont {G.}~\bibnamefont
  {Marques-Tavares}},\ }\bibfield  {title} {\bibinfo {title} {{Relaxation from
  particle production}},\ }\href {https://doi.org/10.1007/JHEP12(2016)101}
  {\bibfield  {journal} {\bibinfo  {journal} {J. High Energy Phys.}\ }\textbf
  {\bibinfo {volume} {2016}}\bibfield  {number} {\bibinfo  {number} { (12)},\
  \bibinfo {pages} {101}},\ }\Eprint {https://arxiv.org/abs/1607.01786}
  {arXiv:1607.01786 [hep-ph]} \BibitemShut {NoStop}%
\bibitem [{\citenamefont {Fonseca}\ \emph
  {et~al.}(2020{\natexlab{a}})\citenamefont {Fonseca}, \citenamefont
  {Morgante}, \citenamefont {Sato},\ and\ \citenamefont
  {Servant}}]{Fonseca:2019ypl}%
  \BibitemOpen
  \bibfield  {author} {\bibinfo {author} {\bibfnamefont {N.}~\bibnamefont
  {Fonseca}}, \bibinfo {author} {\bibfnamefont {E.}~\bibnamefont {Morgante}},
  \bibinfo {author} {\bibfnamefont {R.}~\bibnamefont {Sato}},\ and\ \bibinfo
  {author} {\bibfnamefont {G.}~\bibnamefont {Servant}},\ }\bibfield  {title}
  {\bibinfo {title} {{Axion fragmentation}},\ }\href
  {https://doi.org/10.1007/JHEP04(2020)010} {\bibfield  {journal} {\bibinfo
  {journal} {J. High Energy Phys.}\ }\textbf {\bibinfo {volume}
  {2020}}\bibfield  {number} {\bibinfo  {number} { (04)},\ \bibinfo {pages}
  {010}},\ }\Eprint {https://arxiv.org/abs/1911.08472} {arXiv:1911.08472
  [hep-ph]} \BibitemShut {NoStop}%
\bibitem [{\citenamefont {Fonseca}\ \emph
  {et~al.}(2020{\natexlab{b}})\citenamefont {Fonseca}, \citenamefont
  {Morgante}, \citenamefont {Sato},\ and\ \citenamefont
  {Servant}}]{Fonseca:2019lmc}%
  \BibitemOpen
  \bibfield  {author} {\bibinfo {author} {\bibfnamefont {N.}~\bibnamefont
  {Fonseca}}, \bibinfo {author} {\bibfnamefont {E.}~\bibnamefont {Morgante}},
  \bibinfo {author} {\bibfnamefont {R.}~\bibnamefont {Sato}},\ and\ \bibinfo
  {author} {\bibfnamefont {G.}~\bibnamefont {Servant}},\ }\bibfield  {title}
  {\bibinfo {title} {{Relaxion fluctuations (self-stopping relaxion) and
  overview of relaxion stopping mechanisms}},\ }\href
  {https://doi.org/10.1007/JHEP05(2020)080} {\bibfield  {journal} {\bibinfo
  {journal} {J. High Energy Phys.}\ }\textbf {\bibinfo {volume}
  {2020}}\bibfield  {number} {\bibinfo  {number} { (05)},\ \bibinfo {pages}
  {080}},\ }\bibinfo {note} {[Erratum: JHEP 01, 012 (2021)]},\ \Eprint
  {https://arxiv.org/abs/1911.08473} {arXiv:1911.08473 [hep-ph]} \BibitemShut
  {NoStop}%
\bibitem [{\citenamefont {Wang}(2019)}]{Wang:2018ddr}%
  \BibitemOpen
  \bibfield  {author} {\bibinfo {author} {\bibfnamefont {S.-J.}\ \bibnamefont
  {Wang}},\ }\bibfield  {title} {\bibinfo {title} {{Paper-boat relaxion}},\
  }\href {https://doi.org/10.1103/PhysRevD.99.095026} {\bibfield  {journal}
  {\bibinfo  {journal} {Phys. Rev. D}\ }\textbf {\bibinfo {volume} {99}},\
  \bibinfo {pages} {095026} (\bibinfo {year} {2019})},\ \Eprint
  {https://arxiv.org/abs/1811.06520} {arXiv:1811.06520 [hep-ph]} \BibitemShut
  {NoStop}%
\bibitem [{\citenamefont {Barducci}\ \emph {et~al.}(2021)\citenamefont
  {Barducci}, \citenamefont {Bertuzzo},\ and\ \citenamefont
  {Tupia}}]{Barducci:2020axp}%
  \BibitemOpen
  \bibfield  {author} {\bibinfo {author} {\bibfnamefont {D.}~\bibnamefont
  {Barducci}}, \bibinfo {author} {\bibfnamefont {E.}~\bibnamefont {Bertuzzo}},\
  and\ \bibinfo {author} {\bibfnamefont {M.~A.}\ \bibnamefont {Tupia}},\
  }\bibfield  {title} {\bibinfo {title} {{Gravitational tests of electroweak
  relaxation}},\ }\href {https://doi.org/10.1007/JHEP07(2021)119} {\bibfield
  {journal} {\bibinfo  {journal} {J. High Energy Phys.}\ }\textbf {\bibinfo
  {volume} {2021}}\bibfield  {number} {\bibinfo  {number} { (07)},\ \bibinfo
  {pages} {119}},\ }\Eprint {https://arxiv.org/abs/2011.05795}
  {arXiv:2011.05795 [astro-ph.CO]} \BibitemShut {NoStop}%
\bibitem [{\citenamefont {Khlebnikov}\ and\ \citenamefont
  {Tkachev}(1997)}]{Khlebnikov:1997di}%
  \BibitemOpen
  \bibfield  {author} {\bibinfo {author} {\bibfnamefont {S.~Y.}\ \bibnamefont
  {Khlebnikov}}\ and\ \bibinfo {author} {\bibfnamefont {I.~I.}\ \bibnamefont
  {Tkachev}},\ }\bibfield  {title} {\bibinfo {title} {{Relic gravitational
  waves produced after preheating}},\ }\href
  {https://doi.org/10.1103/PhysRevD.56.653} {\bibfield  {journal} {\bibinfo
  {journal} {Phys. Rev. D}\ }\textbf {\bibinfo {volume} {56}},\ \bibinfo
  {pages} {653} (\bibinfo {year} {1997})},\ \Eprint
  {https://arxiv.org/abs/hep-ph/9701423} {arXiv:hep-ph/9701423} \BibitemShut
  {NoStop}%
\bibitem [{\citenamefont {Easther}\ and\ \citenamefont
  {Lim}(2006)}]{Easther:2006gt}%
  \BibitemOpen
  \bibfield  {author} {\bibinfo {author} {\bibfnamefont {R.}~\bibnamefont
  {Easther}}\ and\ \bibinfo {author} {\bibfnamefont {E.~A.}\ \bibnamefont
  {Lim}},\ }\bibfield  {title} {\bibinfo {title} {{Stochastic gravitational
  wave production after inflation}},\ }\href
  {https://doi.org/10.1088/1475-7516/2006/04/010} {\bibfield  {journal}
  {\bibinfo  {journal} {J. Cosmol. Astropart. Phys.}\ }\textbf {\bibinfo
  {volume} {2006}}\bibfield  {number} {\bibinfo  {number} { (04)},\ \bibinfo
  {pages} {010}},\ }\Eprint {https://arxiv.org/abs/astro-ph/0601617}
  {arXiv:astro-ph/0601617} \BibitemShut {NoStop}%
\bibitem [{\citenamefont {Dufaux}\ \emph {et~al.}(2009)\citenamefont {Dufaux},
  \citenamefont {Felder}, \citenamefont {Kofman},\ and\ \citenamefont
  {Navros}}]{Dufaux:2008dn}%
  \BibitemOpen
  \bibfield  {author} {\bibinfo {author} {\bibfnamefont {J.-F.}\ \bibnamefont
  {Dufaux}}, \bibinfo {author} {\bibfnamefont {G.}~\bibnamefont {Felder}},
  \bibinfo {author} {\bibfnamefont {L.}~\bibnamefont {Kofman}},\ and\ \bibinfo
  {author} {\bibfnamefont {O.}~\bibnamefont {Navros}},\ }\bibfield  {title}
  {\bibinfo {title} {{Gravity waves from tachyonic preheating after hybrid
  inflation}},\ }\href {https://doi.org/10.1088/1475-7516/2009/03/001}
  {\bibfield  {journal} {\bibinfo  {journal} {J. Cosmol. Astropart. Phys.}\
  }\textbf {\bibinfo {volume} {2009}}\bibfield  {number} {\bibinfo  {number} {
  (03)},\ \bibinfo {pages} {001}},\ }\Eprint {https://arxiv.org/abs/0812.2917}
  {arXiv:0812.2917 [astro-ph]} \BibitemShut {NoStop}%
\bibitem [{\citenamefont {Adshead}\ \emph {et~al.}(2018)\citenamefont
  {Adshead}, \citenamefont {Giblin},\ and\ \citenamefont
  {Weiner}}]{Adshead:2018doq}%
  \BibitemOpen
  \bibfield  {author} {\bibinfo {author} {\bibfnamefont {P.}~\bibnamefont
  {Adshead}}, \bibinfo {author} {\bibfnamefont {J.~T.}\ \bibnamefont
  {Giblin}},\ and\ \bibinfo {author} {\bibfnamefont {Z.~J.}\ \bibnamefont
  {Weiner}},\ }\bibfield  {title} {\bibinfo {title} {{Gravitational waves from
  gauge preheating}},\ }\href {https://doi.org/10.1103/PhysRevD.98.043525}
  {\bibfield  {journal} {\bibinfo  {journal} {Phys. Rev. D}\ }\textbf {\bibinfo
  {volume} {98}},\ \bibinfo {pages} {043525} (\bibinfo {year} {2018})},\
  \Eprint {https://arxiv.org/abs/1805.04550} {arXiv:1805.04550 [astro-ph.CO]}
  \BibitemShut {NoStop}%
\bibitem [{\citenamefont {Machado}\ \emph {et~al.}(2019)\citenamefont
  {Machado}, \citenamefont {Ratzinger}, \citenamefont {Schwaller},\ and\
  \citenamefont {Stefanek}}]{Machado:2018nqk}%
  \BibitemOpen
  \bibfield  {author} {\bibinfo {author} {\bibfnamefont {C.~S.}\ \bibnamefont
  {Machado}}, \bibinfo {author} {\bibfnamefont {W.}~\bibnamefont {Ratzinger}},
  \bibinfo {author} {\bibfnamefont {P.}~\bibnamefont {Schwaller}},\ and\
  \bibinfo {author} {\bibfnamefont {B.~A.}\ \bibnamefont {Stefanek}},\
  }\bibfield  {title} {\bibinfo {title} {{Audible axions}},\ }\href
  {https://doi.org/10.1007/JHEP01(2019)053} {\bibfield  {journal} {\bibinfo
  {journal} {J. High Energy Phys.}\ }\textbf {\bibinfo {volume}
  {2019}}\bibfield  {number} {\bibinfo  {number} { (01)},\ \bibinfo {pages}
  {053}},\ }\Eprint {https://arxiv.org/abs/1811.01950} {arXiv:1811.01950
  [hep-ph]} \BibitemShut {NoStop}%
\bibitem [{\citenamefont {Machado}\ \emph {et~al.}(2020)\citenamefont
  {Machado}, \citenamefont {Ratzinger}, \citenamefont {Schwaller},\ and\
  \citenamefont {Stefanek}}]{Machado:2019xuc}%
  \BibitemOpen
  \bibfield  {author} {\bibinfo {author} {\bibfnamefont {C.~S.}\ \bibnamefont
  {Machado}}, \bibinfo {author} {\bibfnamefont {W.}~\bibnamefont {Ratzinger}},
  \bibinfo {author} {\bibfnamefont {P.}~\bibnamefont {Schwaller}},\ and\
  \bibinfo {author} {\bibfnamefont {B.~A.}\ \bibnamefont {Stefanek}},\
  }\bibfield  {title} {\bibinfo {title} {{Gravitational wave probes of
  axionlike particles}},\ }\href {https://doi.org/10.1103/PhysRevD.102.075033}
  {\bibfield  {journal} {\bibinfo  {journal} {Phys. Rev. D}\ }\textbf {\bibinfo
  {volume} {102}},\ \bibinfo {pages} {075033} (\bibinfo {year} {2020})},\
  \Eprint {https://arxiv.org/abs/1912.01007} {arXiv:1912.01007 [hep-ph]}
  \BibitemShut {NoStop}%
\bibitem [{\citenamefont {Ratzinger}\ \emph {et~al.}(2021)\citenamefont
  {Ratzinger}, \citenamefont {Schwaller},\ and\ \citenamefont
  {Stefanek}}]{Ratzinger:2020oct}%
  \BibitemOpen
  \bibfield  {author} {\bibinfo {author} {\bibfnamefont {W.}~\bibnamefont
  {Ratzinger}}, \bibinfo {author} {\bibfnamefont {P.}~\bibnamefont
  {Schwaller}},\ and\ \bibinfo {author} {\bibfnamefont {B.~A.}\ \bibnamefont
  {Stefanek}},\ }\bibfield  {title} {\bibinfo {title} {{Gravitational waves
  from an axion-dark photon system: A lattice study}},\ }\href
  {https://doi.org/10.21468/SciPostPhys.11.1.001} {\bibfield  {journal}
  {\bibinfo  {journal} {SciPost Phys.}\ }\textbf {\bibinfo {volume} {11}},\
  \bibinfo {pages} {001} (\bibinfo {year} {2021})},\ \Eprint
  {https://arxiv.org/abs/2012.11584} {arXiv:2012.11584 [astro-ph.CO]}
  \BibitemShut {NoStop}%
\bibitem [{\citenamefont {Choi}\ \emph {et~al.}(2014)\citenamefont {Choi},
  \citenamefont {Kim},\ and\ \citenamefont {Yun}}]{Choi:2014rja}%
  \BibitemOpen
  \bibfield  {author} {\bibinfo {author} {\bibfnamefont {K.}~\bibnamefont
  {Choi}}, \bibinfo {author} {\bibfnamefont {H.}~\bibnamefont {Kim}},\ and\
  \bibinfo {author} {\bibfnamefont {S.}~\bibnamefont {Yun}},\ }\bibfield
  {title} {\bibinfo {title} {{Natural inflation with multiple sub-Planckian
  axions}},\ }\href {https://doi.org/10.1103/PhysRevD.90.023545} {\bibfield
  {journal} {\bibinfo  {journal} {Phys. Rev. D}\ }\textbf {\bibinfo {volume}
  {90}},\ \bibinfo {pages} {023545} (\bibinfo {year} {2014})},\ \Eprint
  {https://arxiv.org/abs/1404.6209} {arXiv:1404.6209 [hep-th]} \BibitemShut
  {NoStop}%
\bibitem [{\citenamefont {Choi}\ and\ \citenamefont
  {Im}(2016{\natexlab{b}})}]{Choi:2015fiu}%
  \BibitemOpen
  \bibfield  {author} {\bibinfo {author} {\bibfnamefont {K.}~\bibnamefont
  {Choi}}\ and\ \bibinfo {author} {\bibfnamefont {S.~H.}\ \bibnamefont {Im}},\
  }\bibfield  {title} {\bibinfo {title} {{Realizing the relaxion from multiple
  axions and its UV completion with high scale supersymmetry}},\ }\href
  {https://doi.org/10.1007/JHEP01(2016)149} {\bibfield  {journal} {\bibinfo
  {journal} {J. High Energy Phys.}\ }\textbf {\bibinfo {volume}
  {2016}}\bibfield  {number} {\bibinfo  {number} { (01)},\ \bibinfo {pages}
  {149}},\ }\Eprint {https://arxiv.org/abs/1511.00132} {arXiv:1511.00132
  [hep-ph]} \BibitemShut {NoStop}%
\bibitem [{\citenamefont {Kaplan}\ and\ \citenamefont
  {Rattazzi}(2016)}]{Kaplan:2015fuy}%
  \BibitemOpen
  \bibfield  {author} {\bibinfo {author} {\bibfnamefont {D.~E.}\ \bibnamefont
  {Kaplan}}\ and\ \bibinfo {author} {\bibfnamefont {R.}~\bibnamefont
  {Rattazzi}},\ }\bibfield  {title} {\bibinfo {title} {{Large field excursions
  and approximate discrete symmetries from a clockwork axion}},\ }\href
  {https://doi.org/10.1103/PhysRevD.93.085007} {\bibfield  {journal} {\bibinfo
  {journal} {Phys. Rev. D}\ }\textbf {\bibinfo {volume} {93}},\ \bibinfo
  {pages} {085007} (\bibinfo {year} {2016})},\ \Eprint
  {https://arxiv.org/abs/1511.01827} {arXiv:1511.01827 [hep-ph]} \BibitemShut
  {NoStop}%
\bibitem [{\citenamefont {Giudice}\ and\ \citenamefont
  {McCullough}(2017)}]{Giudice:2016yja}%
  \BibitemOpen
  \bibfield  {author} {\bibinfo {author} {\bibfnamefont {G.~F.}\ \bibnamefont
  {Giudice}}\ and\ \bibinfo {author} {\bibfnamefont {M.}~\bibnamefont
  {McCullough}},\ }\bibfield  {title} {\bibinfo {title} {{A clockwork
  theory}},\ }\href {https://doi.org/10.1007/JHEP02(2017)036} {\bibfield
  {journal} {\bibinfo  {journal} {J. High Energy Phys.}\ }\textbf {\bibinfo
  {volume} {2017}}\bibfield  {number} {\bibinfo  {number} { (02)},\ \bibinfo
  {pages} {036}},\ }\Eprint {https://arxiv.org/abs/1610.07962}
  {arXiv:1610.07962 [hep-ph]} \BibitemShut {NoStop}%
\bibitem [{\citenamefont {Aghanim}\ \emph {et~al.}(2020)\citenamefont {Aghanim}
  \emph {et~al.}}]{Aghanim:2018eyx}%
  \BibitemOpen
  \bibfield  {author} {\bibinfo {author} {\bibfnamefont {N.}~\bibnamefont
  {Aghanim}} \emph {et~al.} (\bibinfo {collaboration} {Planck}),\ }\bibfield
  {title} {\bibinfo {title} {{Planck 2018 results. VI. Cosmological
  parameters}},\ }\href {https://doi.org/10.1051/0004-6361/201833910}
  {\bibfield  {journal} {\bibinfo  {journal} {Astron. Astrophys.}\ }\textbf
  {\bibinfo {volume} {641}},\ \bibinfo {pages} {A6} (\bibinfo {year} {2020})},\
  \Eprint {https://arxiv.org/abs/1807.06209} {arXiv:1807.06209 [astro-ph.CO]}
  \BibitemShut {NoStop}%
\bibitem [{\citenamefont {Kolb}\ and\ \citenamefont
  {Turner}(1990)}]{Kolb:1990vq}%
  \BibitemOpen
  \bibfield  {author} {\bibinfo {author} {\bibfnamefont {E.~W.}\ \bibnamefont
  {Kolb}}\ and\ \bibinfo {author} {\bibfnamefont {M.~S.}\ \bibnamefont
  {Turner}},\ }\href@noop {} {\emph {\bibinfo {title} {{The Early
  Universe}}}},\ \bibinfo {series} {{Frontiers in Physics}}, Vol.~\bibinfo
  {volume} {69}\ (\bibinfo  {publisher} {Westview Press},\ \bibinfo {year}
  {1990})\BibitemShut {NoStop}%
\bibitem [{\citenamefont {Audren}\ \emph {et~al.}(2014)\citenamefont {Audren},
  \citenamefont {Lesgourgues}, \citenamefont {Mangano}, \citenamefont
  {Serpico},\ and\ \citenamefont {Tram}}]{Audren:2014bca}%
  \BibitemOpen
  \bibfield  {author} {\bibinfo {author} {\bibfnamefont {B.}~\bibnamefont
  {Audren}}, \bibinfo {author} {\bibfnamefont {J.}~\bibnamefont {Lesgourgues}},
  \bibinfo {author} {\bibfnamefont {G.}~\bibnamefont {Mangano}}, \bibinfo
  {author} {\bibfnamefont {P.~D.}\ \bibnamefont {Serpico}},\ and\ \bibinfo
  {author} {\bibfnamefont {T.}~\bibnamefont {Tram}},\ }\bibfield  {title}
  {\bibinfo {title} {{Strongest model-independent bound on the lifetime of Dark
  Matter}},\ }\href {https://doi.org/10.1088/1475-7516/2014/12/028} {\bibfield
  {journal} {\bibinfo  {journal} {J. Cosmol. Astropart. Phys.}\ }\textbf
  {\bibinfo {volume} {2014}}\bibfield  {number} {\bibinfo  {number} { (12)},\
  \bibinfo {pages} {028}},\ }\Eprint {https://arxiv.org/abs/1407.2418}
  {arXiv:1407.2418 [astro-ph.CO]} \BibitemShut {NoStop}%
\bibitem [{\citenamefont {Agrawal}\ \emph {et~al.}(2018)\citenamefont
  {Agrawal}, \citenamefont {Marques-Tavares},\ and\ \citenamefont
  {Xue}}]{Agrawal:2017eqm}%
  \BibitemOpen
  \bibfield  {author} {\bibinfo {author} {\bibfnamefont {P.}~\bibnamefont
  {Agrawal}}, \bibinfo {author} {\bibfnamefont {G.}~\bibnamefont
  {Marques-Tavares}},\ and\ \bibinfo {author} {\bibfnamefont {W.}~\bibnamefont
  {Xue}},\ }\bibfield  {title} {\bibinfo {title} {{Opening up the QCD axion
  window}},\ }\href {https://doi.org/10.1007/JHEP03(2018)049} {\bibfield
  {journal} {\bibinfo  {journal} {J. High Energy Phys.}\ }\textbf {\bibinfo
  {volume} {2018}}\bibfield  {number} {\bibinfo  {number} { (03)},\ \bibinfo
  {pages} {049}},\ }\Eprint {https://arxiv.org/abs/1708.05008}
  {arXiv:1708.05008 [hep-ph]} \BibitemShut {NoStop}%
\bibitem [{\citenamefont {Kitajima}\ \emph
  {et~al.}(2018{\natexlab{a}})\citenamefont {Kitajima}, \citenamefont
  {Sekiguchi},\ and\ \citenamefont {Takahashi}}]{Kitajima:2017peg}%
  \BibitemOpen
  \bibfield  {author} {\bibinfo {author} {\bibfnamefont {N.}~\bibnamefont
  {Kitajima}}, \bibinfo {author} {\bibfnamefont {T.}~\bibnamefont
  {Sekiguchi}},\ and\ \bibinfo {author} {\bibfnamefont {F.}~\bibnamefont
  {Takahashi}},\ }\bibfield  {title} {\bibinfo {title} {{Cosmological abundance
  of the QCD axion coupled to hidden photons}},\ }\href
  {https://doi.org/10.1016/j.physletb.2018.04.024} {\bibfield  {journal}
  {\bibinfo  {journal} {Phys. Lett. B}\ }\textbf {\bibinfo {volume} {781}},\
  \bibinfo {pages} {684} (\bibinfo {year} {2018}{\natexlab{a}})},\ \Eprint
  {https://arxiv.org/abs/1711.06590} {arXiv:1711.06590 [hep-ph]} \BibitemShut
  {NoStop}%
\bibitem [{\citenamefont {Kitajima}\ \emph {et~al.}(2021)\citenamefont
  {Kitajima}, \citenamefont {Soda},\ and\ \citenamefont
  {Urakawa}}]{Kitajima:2020rpm}%
  \BibitemOpen
  \bibfield  {author} {\bibinfo {author} {\bibfnamefont {N.}~\bibnamefont
  {Kitajima}}, \bibinfo {author} {\bibfnamefont {J.}~\bibnamefont {Soda}},\
  and\ \bibinfo {author} {\bibfnamefont {Y.}~\bibnamefont {Urakawa}},\
  }\bibfield  {title} {\bibinfo {title} {{Nano-Hz Gravitational-Wave Signature
  from Axion Dark Matter}},\ }\href
  {https://doi.org/10.1103/PhysRevLett.126.121301} {\bibfield  {journal}
  {\bibinfo  {journal} {Phys. Rev. Lett.}\ }\textbf {\bibinfo {volume} {126}},\
  \bibinfo {pages} {121301} (\bibinfo {year} {2021})},\ \Eprint
  {https://arxiv.org/abs/2010.10990} {arXiv:2010.10990 [astro-ph.CO]}
  \BibitemShut {NoStop}%
\bibitem [{\citenamefont {O'Hare}\ and\ \citenamefont
  {Vitagliano}(2020)}]{OHare:2020wah}%
  \BibitemOpen
  \bibfield  {author} {\bibinfo {author} {\bibfnamefont {C.~A.~J.}\
  \bibnamefont {O'Hare}}\ and\ \bibinfo {author} {\bibfnamefont
  {E.}~\bibnamefont {Vitagliano}},\ }\bibfield  {title} {\bibinfo {title}
  {{Cornering the axion with $CP$-violating interactions}},\ }\href
  {https://doi.org/10.1103/PhysRevD.102.115026} {\bibfield  {journal} {\bibinfo
   {journal} {Phys. Rev. D}\ }\textbf {\bibinfo {volume} {102}},\ \bibinfo
  {pages} {115026} (\bibinfo {year} {2020})},\ \Eprint
  {https://arxiv.org/abs/2010.03889} {arXiv:2010.03889 [hep-ph]} \BibitemShut
  {NoStop}%
\bibitem [{\citenamefont {Tan}\ \emph {et~al.}(2020)\citenamefont {Tan} \emph
  {et~al.}}]{Tan:2020vpf}%
  \BibitemOpen
  \bibfield  {author} {\bibinfo {author} {\bibfnamefont {W.-H.}\ \bibnamefont
  {Tan}} \emph {et~al.},\ }\bibfield  {title} {\bibinfo {title} {{Improvement
  for Testing the Gravitational Inverse-Square Law at the Submillimeter
  Range}},\ }\href {https://doi.org/10.1103/PhysRevLett.124.051301} {\bibfield
  {journal} {\bibinfo  {journal} {Phys. Rev. Lett.}\ }\textbf {\bibinfo
  {volume} {124}},\ \bibinfo {pages} {051301} (\bibinfo {year}
  {2020})}\BibitemShut {NoStop}%
\bibitem [{\citenamefont {Berg\'e}\ \emph {et~al.}(2018)\citenamefont
  {Berg\'e}, \citenamefont {Brax}, \citenamefont {M\'etris}, \citenamefont
  {Pernot-Borr\`as}, \citenamefont {Touboul},\ and\ \citenamefont
  {Uzan}}]{Berge:2017ovy}%
  \BibitemOpen
  \bibfield  {author} {\bibinfo {author} {\bibfnamefont {J.}~\bibnamefont
  {Berg\'e}}, \bibinfo {author} {\bibfnamefont {P.}~\bibnamefont {Brax}},
  \bibinfo {author} {\bibfnamefont {G.}~\bibnamefont {M\'etris}}, \bibinfo
  {author} {\bibfnamefont {M.}~\bibnamefont {Pernot-Borr\`as}}, \bibinfo
  {author} {\bibfnamefont {P.}~\bibnamefont {Touboul}},\ and\ \bibinfo {author}
  {\bibfnamefont {J.-P.}\ \bibnamefont {Uzan}},\ }\bibfield  {title} {\bibinfo
  {title} {{MICROSCOPE Mission: First Constraints on the Violation of the Weak
  Equivalence Principle by a Light Scalar Dilaton}},\ }\href
  {https://doi.org/10.1103/PhysRevLett.120.141101} {\bibfield  {journal}
  {\bibinfo  {journal} {Phys. Rev. Lett.}\ }\textbf {\bibinfo {volume} {120}},\
  \bibinfo {pages} {141101} (\bibinfo {year} {2018})},\ \Eprint
  {https://arxiv.org/abs/1712.00483} {arXiv:1712.00483 [gr-qc]} \BibitemShut
  {NoStop}%
\bibitem [{\citenamefont {Chen}\ \emph {et~al.}(2016)\citenamefont {Chen},
  \citenamefont {Tham}, \citenamefont {Krause}, \citenamefont {L\'{o}pez},
  \citenamefont {Fischbach},\ and\ \citenamefont {Decca}}]{Chen:2014oda}%
  \BibitemOpen
  \bibfield  {author} {\bibinfo {author} {\bibfnamefont {Y.~J.}\ \bibnamefont
  {Chen}}, \bibinfo {author} {\bibfnamefont {W.~K.}\ \bibnamefont {Tham}},
  \bibinfo {author} {\bibfnamefont {D.~E.}\ \bibnamefont {Krause}}, \bibinfo
  {author} {\bibfnamefont {D.}~\bibnamefont {L\'{o}pez}}, \bibinfo {author}
  {\bibfnamefont {E.}~\bibnamefont {Fischbach}},\ and\ \bibinfo {author}
  {\bibfnamefont {R.~S.}\ \bibnamefont {Decca}},\ }\bibfield  {title} {\bibinfo
  {title} {{Stronger Limits on Hypothetical Yukawa Interactions in the
  30\textendash{}8000 nm Range}},\ }\href
  {https://doi.org/10.1103/PhysRevLett.116.221102} {\bibfield  {journal}
  {\bibinfo  {journal} {Phys. Rev. Lett.}\ }\textbf {\bibinfo {volume} {116}},\
  \bibinfo {pages} {221102} (\bibinfo {year} {2016})},\ \Eprint
  {https://arxiv.org/abs/1410.7267} {arXiv:1410.7267 [hep-ex]} \BibitemShut
  {NoStop}%
\bibitem [{\citenamefont {Kapner}\ \emph {et~al.}(2007)\citenamefont {Kapner},
  \citenamefont {Cook}, \citenamefont {Adelberger}, \citenamefont {Gundlach},
  \citenamefont {Heckel}, \citenamefont {Hoyle},\ and\ \citenamefont
  {Swanson}}]{Kapner:2006si}%
  \BibitemOpen
  \bibfield  {author} {\bibinfo {author} {\bibfnamefont {D.~J.}\ \bibnamefont
  {Kapner}}, \bibinfo {author} {\bibfnamefont {T.~S.}\ \bibnamefont {Cook}},
  \bibinfo {author} {\bibfnamefont {E.~G.}\ \bibnamefont {Adelberger}},
  \bibinfo {author} {\bibfnamefont {J.~H.}\ \bibnamefont {Gundlach}}, \bibinfo
  {author} {\bibfnamefont {B.~R.}\ \bibnamefont {Heckel}}, \bibinfo {author}
  {\bibfnamefont {C.~D.}\ \bibnamefont {Hoyle}},\ and\ \bibinfo {author}
  {\bibfnamefont {H.~E.}\ \bibnamefont {Swanson}},\ }\bibfield  {title}
  {\bibinfo {title} {{Tests of the Gravitational Inverse-Square Law Below the
  Dark-Energy Length Scale}},\ }\href
  {https://doi.org/10.1103/PhysRevLett.98.021101} {\bibfield  {journal}
  {\bibinfo  {journal} {Phys. Rev. Lett.}\ }\textbf {\bibinfo {volume} {98}},\
  \bibinfo {pages} {021101} (\bibinfo {year} {2007})},\ \Eprint
  {https://arxiv.org/abs/hep-ph/0611184} {arXiv:hep-ph/0611184} \BibitemShut
  {NoStop}%
\bibitem [{\citenamefont {Schlamminger}\ \emph {et~al.}(2008)\citenamefont
  {Schlamminger}, \citenamefont {Choi}, \citenamefont {Wagner}, \citenamefont
  {Gundlach},\ and\ \citenamefont {Adelberger}}]{Schlamminger:2007ht}%
  \BibitemOpen
  \bibfield  {author} {\bibinfo {author} {\bibfnamefont {S.}~\bibnamefont
  {Schlamminger}}, \bibinfo {author} {\bibfnamefont {K.~Y.}\ \bibnamefont
  {Choi}}, \bibinfo {author} {\bibfnamefont {T.~A.}\ \bibnamefont {Wagner}},
  \bibinfo {author} {\bibfnamefont {J.~H.}\ \bibnamefont {Gundlach}},\ and\
  \bibinfo {author} {\bibfnamefont {E.~G.}\ \bibnamefont {Adelberger}},\
  }\bibfield  {title} {\bibinfo {title} {{Test of the Equivalence Principle
  Using a Rotating Torsion Balance}},\ }\href
  {https://doi.org/10.1103/PhysRevLett.100.041101} {\bibfield  {journal}
  {\bibinfo  {journal} {Phys. Rev. Lett.}\ }\textbf {\bibinfo {volume} {100}},\
  \bibinfo {pages} {041101} (\bibinfo {year} {2008})},\ \Eprint
  {https://arxiv.org/abs/0712.0607} {arXiv:0712.0607 [gr-qc]} \BibitemShut
  {NoStop}%
\bibitem [{\citenamefont {Hoskins}\ \emph {et~al.}(1985)\citenamefont
  {Hoskins}, \citenamefont {Newman}, \citenamefont {Spero},\ and\ \citenamefont
  {Schultz}}]{Hoskins:1985tn}%
  \BibitemOpen
  \bibfield  {author} {\bibinfo {author} {\bibfnamefont {J.~K.}\ \bibnamefont
  {Hoskins}}, \bibinfo {author} {\bibfnamefont {R.~D.}\ \bibnamefont {Newman}},
  \bibinfo {author} {\bibfnamefont {R.}~\bibnamefont {Spero}},\ and\ \bibinfo
  {author} {\bibfnamefont {J.}~\bibnamefont {Schultz}},\ }\bibfield  {title}
  {\bibinfo {title} {{Experimental tests of the gravitational inverse square
  law for mass separations from 2 to 105~cm}},\ }\href
  {https://doi.org/10.1103/PhysRevD.32.3084} {\bibfield  {journal} {\bibinfo
  {journal} {Phys. Rev. D}\ }\textbf {\bibinfo {volume} {32}},\ \bibinfo
  {pages} {3084} (\bibinfo {year} {1985})}\BibitemShut {NoStop}%
\bibitem [{\citenamefont {Grifols}\ \emph {et~al.}(1989)\citenamefont
  {Grifols}, \citenamefont {Masso},\ and\ \citenamefont
  {Peris}}]{Grifols:1988fv}%
  \BibitemOpen
  \bibfield  {author} {\bibinfo {author} {\bibfnamefont {J.~A.}\ \bibnamefont
  {Grifols}}, \bibinfo {author} {\bibfnamefont {E.}~\bibnamefont {Masso}},\
  and\ \bibinfo {author} {\bibfnamefont {S.}~\bibnamefont {Peris}},\ }\bibfield
   {title} {\bibinfo {title} {{Energy loss From the sun and {RED} giants:
  Bounds on short range baryonic and leptonic forces}},\ }\href
  {https://doi.org/10.1142/S0217732389000381} {\bibfield  {journal} {\bibinfo
  {journal} {Mod. Phys. Lett. A}\ }\textbf {\bibinfo {volume} {4}},\ \bibinfo
  {pages} {311} (\bibinfo {year} {1989})}\BibitemShut {NoStop}%
\bibitem [{\citenamefont {Cadamuro}\ and\ \citenamefont
  {Redondo}(2012)}]{Cadamuro:2011fd}%
  \BibitemOpen
  \bibfield  {author} {\bibinfo {author} {\bibfnamefont {D.}~\bibnamefont
  {Cadamuro}}\ and\ \bibinfo {author} {\bibfnamefont {J.}~\bibnamefont
  {Redondo}},\ }\bibfield  {title} {\bibinfo {title} {{Cosmological bounds on
  pseudo Nambu-Goldstone bosons}},\ }\href
  {https://doi.org/10.1088/1475-7516/2012/02/032} {\bibfield  {journal}
  {\bibinfo  {journal} {J. Cosmol. Astropart. Phys.}\ }\textbf {\bibinfo
  {volume} {2012}}\bibfield  {number} {\bibinfo  {number} { (02)},\ \bibinfo
  {pages} {032}},\ }\Eprint {https://arxiv.org/abs/1110.2895} {arXiv:1110.2895
  [hep-ph]} \BibitemShut {NoStop}%
\bibitem [{\citenamefont {Raffelt}(2012)}]{Raffelt:2012sp}%
  \BibitemOpen
  \bibfield  {author} {\bibinfo {author} {\bibfnamefont {G.}~\bibnamefont
  {Raffelt}},\ }\bibfield  {title} {\bibinfo {title} {{Limits on a
  $CP$-violating scalar axion-nucleon interaction}},\ }\href
  {https://doi.org/10.1103/PhysRevD.86.015001} {\bibfield  {journal} {\bibinfo
  {journal} {Phys. Rev. D}\ }\textbf {\bibinfo {volume} {86}},\ \bibinfo
  {pages} {015001} (\bibinfo {year} {2012})},\ \Eprint
  {https://arxiv.org/abs/1205.1776} {arXiv:1205.1776 [hep-ph]} \BibitemShut
  {NoStop}%
\bibitem [{\citenamefont {Hardy}\ and\ \citenamefont
  {Lasenby}(2017)}]{Hardy:2016kme}%
  \BibitemOpen
  \bibfield  {author} {\bibinfo {author} {\bibfnamefont {E.}~\bibnamefont
  {Hardy}}\ and\ \bibinfo {author} {\bibfnamefont {R.}~\bibnamefont
  {Lasenby}},\ }\bibfield  {title} {\bibinfo {title} {{Stellar cooling bounds
  on new light particles: plasma mixing effects}},\ }\href
  {https://doi.org/10.1007/JHEP02(2017)033} {\bibfield  {journal} {\bibinfo
  {journal} {J. High Energy Phys.}\ }\textbf {\bibinfo {volume}
  {2017}}\bibfield  {number} {\bibinfo  {number} { (02)},\ \bibinfo {pages}
  {033}},\ }\Eprint {https://arxiv.org/abs/1611.05852} {arXiv:1611.05852
  [hep-ph]} \BibitemShut {NoStop}%
\bibitem [{\citenamefont {Budnik}\ \emph {et~al.}(2019)\citenamefont {Budnik},
  \citenamefont {Davidi}, \citenamefont {Kim}, \citenamefont {Perez},\ and\
  \citenamefont {Priel}}]{Budnik:2019olh}%
  \BibitemOpen
  \bibfield  {author} {\bibinfo {author} {\bibfnamefont {R.}~\bibnamefont
  {Budnik}}, \bibinfo {author} {\bibfnamefont {O.}~\bibnamefont {Davidi}},
  \bibinfo {author} {\bibfnamefont {H.}~\bibnamefont {Kim}}, \bibinfo {author}
  {\bibfnamefont {G.}~\bibnamefont {Perez}},\ and\ \bibinfo {author}
  {\bibfnamefont {N.}~\bibnamefont {Priel}},\ }\bibfield  {title} {\bibinfo
  {title} {{Searching for a solar relaxion or scalar particle with XENON1T and
  LUX}},\ }\href {https://doi.org/10.1103/PhysRevD.100.095021} {\bibfield
  {journal} {\bibinfo  {journal} {Phys. Rev. D}\ }\textbf {\bibinfo {volume}
  {100}},\ \bibinfo {pages} {095021} (\bibinfo {year} {2019})},\ \Eprint
  {https://arxiv.org/abs/1909.02568} {arXiv:1909.02568 [hep-ph]} \BibitemShut
  {NoStop}%
\bibitem [{\citenamefont {Riess}\ \emph {et~al.}(2018)\citenamefont {Riess}
  \emph {et~al.}}]{Riess:2018uxu}%
  \BibitemOpen
  \bibfield  {author} {\bibinfo {author} {\bibfnamefont {A.~G.}\ \bibnamefont
  {Riess}} \emph {et~al.},\ }\bibfield  {title} {\bibinfo {title} {{New
  parallaxes of galactic cepheids from spatially scanning the Hubble space
  telescope: Implications for the Hubble constant}},\ }\href
  {https://doi.org/10.3847/1538-4357/aaadb7} {\bibfield  {journal} {\bibinfo
  {journal} {Astrophys. J.}\ }\textbf {\bibinfo {volume} {855}},\ \bibinfo
  {pages} {136} (\bibinfo {year} {2018})},\ \Eprint
  {https://arxiv.org/abs/1801.01120} {arXiv:1801.01120 [astro-ph.SR]}
  \BibitemShut {NoStop}%
\bibitem [{\citenamefont {Salehian}\ \emph {et~al.}(2021)\citenamefont
  {Salehian}, \citenamefont {Gorji}, \citenamefont {Mukohyama},\ and\
  \citenamefont {Firouzjahi}}]{Salehian:2020dsf}%
  \BibitemOpen
  \bibfield  {author} {\bibinfo {author} {\bibfnamefont {B.}~\bibnamefont
  {Salehian}}, \bibinfo {author} {\bibfnamefont {M.~A.}\ \bibnamefont {Gorji}},
  \bibinfo {author} {\bibfnamefont {S.}~\bibnamefont {Mukohyama}},\ and\
  \bibinfo {author} {\bibfnamefont {H.}~\bibnamefont {Firouzjahi}},\ }\bibfield
   {title} {\bibinfo {title} {{Analytic study of dark photon and gravitational
  wave production from axion}},\ }\href
  {https://doi.org/10.1007/JHEP05(2021)043} {\bibfield  {journal} {\bibinfo
  {journal} {J. High Energy Phys.}\ }\textbf {\bibinfo {volume}
  {2021}}\bibfield  {number} {\bibinfo  {number} { (05)},\ \bibinfo {pages}
  {043}},\ }\Eprint {https://arxiv.org/abs/2007.08148} {arXiv:2007.08148
  [hep-ph]} \BibitemShut {NoStop}%
\bibitem [{\citenamefont {Soda}\ and\ \citenamefont
  {Urakawa}(2018)}]{Soda:2017dsu}%
  \BibitemOpen
  \bibfield  {author} {\bibinfo {author} {\bibfnamefont {J.}~\bibnamefont
  {Soda}}\ and\ \bibinfo {author} {\bibfnamefont {Y.}~\bibnamefont {Urakawa}},\
  }\bibfield  {title} {\bibinfo {title} {{Cosmological imprints of string
  axions in plateau}},\ }\href {https://doi.org/10.1140/epjc/s10052-018-6246-6}
  {\bibfield  {journal} {\bibinfo  {journal} {Eur. Phys. J. C}\ }\textbf
  {\bibinfo {volume} {78}},\ \bibinfo {pages} {779} (\bibinfo {year} {2018})},\
  \Eprint {https://arxiv.org/abs/1710.00305} {arXiv:1710.00305 [astro-ph.CO]}
  \BibitemShut {NoStop}%
\bibitem [{\citenamefont {Kitajima}\ \emph
  {et~al.}(2018{\natexlab{b}})\citenamefont {Kitajima}, \citenamefont {Soda},\
  and\ \citenamefont {Urakawa}}]{Kitajima:2018zco}%
  \BibitemOpen
  \bibfield  {author} {\bibinfo {author} {\bibfnamefont {N.}~\bibnamefont
  {Kitajima}}, \bibinfo {author} {\bibfnamefont {J.}~\bibnamefont {Soda}},\
  and\ \bibinfo {author} {\bibfnamefont {Y.}~\bibnamefont {Urakawa}},\
  }\bibfield  {title} {\bibinfo {title} {{Gravitational wave forest from string
  axiverse}},\ }\href {https://doi.org/10.1088/1475-7516/2018/10/008}
  {\bibfield  {journal} {\bibinfo  {journal} {J. Cosmol. Astropart. Phys.}\
  }\textbf {\bibinfo {volume} {2018}}\bibfield  {number} {\bibinfo  {number} {
  (10)},\ \bibinfo {pages} {008}},\ }\Eprint {https://arxiv.org/abs/1807.07037}
  {arXiv:1807.07037 [astro-ph.CO]} \BibitemShut {NoStop}%
\bibitem [{\citenamefont {Chatrchyan}\ and\ \citenamefont
  {Jaeckel}(2021)}]{Chatrchyan:2020pzh}%
  \BibitemOpen
  \bibfield  {author} {\bibinfo {author} {\bibfnamefont {A.}~\bibnamefont
  {Chatrchyan}}\ and\ \bibinfo {author} {\bibfnamefont {J.}~\bibnamefont
  {Jaeckel}},\ }\bibfield  {title} {\bibinfo {title} {{Gravitational waves from
  the fragmentation of axion-like particle dark matter}},\ }\href
  {https://doi.org/10.1088/1475-7516/2021/02/003} {\bibfield  {journal}
  {\bibinfo  {journal} {J. Cosmol. Astropart. Phys.}\ }\textbf {\bibinfo
  {volume} {2021}}\bibfield  {number} {\bibinfo  {number} { (02)},\ \bibinfo
  {pages} {003}},\ }\Eprint {https://arxiv.org/abs/2004.07844}
  {arXiv:2004.07844 [hep-ph]} \BibitemShut {NoStop}%
\bibitem [{\citenamefont {Thrane}\ and\ \citenamefont
  {Romano}(2013)}]{Thrane:2013oya}%
  \BibitemOpen
  \bibfield  {author} {\bibinfo {author} {\bibfnamefont {E.}~\bibnamefont
  {Thrane}}\ and\ \bibinfo {author} {\bibfnamefont {J.~D.}\ \bibnamefont
  {Romano}},\ }\bibfield  {title} {\bibinfo {title} {{Sensitivity curves for
  searches for gravitational-wave backgrounds}},\ }\href
  {https://doi.org/10.1103/PhysRevD.88.124032} {\bibfield  {journal} {\bibinfo
  {journal} {Phys. Rev. D}\ }\textbf {\bibinfo {volume} {88}},\ \bibinfo
  {pages} {124032} (\bibinfo {year} {2013})},\ \Eprint
  {https://arxiv.org/abs/1310.5300} {arXiv:1310.5300 [astro-ph.IM]}
  \BibitemShut {NoStop}%
\bibitem [{\citenamefont {Amaro-Seoane}\ \emph {et~al.}(2017)\citenamefont
  {Amaro-Seoane} \emph {et~al.}}]{Audley:2017drz}%
  \BibitemOpen
  \bibfield  {author} {\bibinfo {author} {\bibfnamefont {P.}~\bibnamefont
  {Amaro-Seoane}} \emph {et~al.} (\bibinfo {collaboration} {LISA}),\ }\bibfield
   {title} {\bibinfo {title} {{Laser Interferometer Space Antenna}},\
  }\href@noop {} {\  (\bibinfo {year} {2017})},\ \Eprint
  {https://arxiv.org/abs/1702.00786} {arXiv:1702.00786 [astro-ph.IM]}
  \BibitemShut {NoStop}%
\bibitem [{\citenamefont {Robson}\ \emph {et~al.}(2019)\citenamefont {Robson},
  \citenamefont {Cornish},\ and\ \citenamefont {Liu}}]{Cornish:2018dyw}%
  \BibitemOpen
  \bibfield  {author} {\bibinfo {author} {\bibfnamefont {T.}~\bibnamefont
  {Robson}}, \bibinfo {author} {\bibfnamefont {N.~J.}\ \bibnamefont
  {Cornish}},\ and\ \bibinfo {author} {\bibfnamefont {C.}~\bibnamefont {Liu}},\
  }\bibfield  {title} {\bibinfo {title} {{The construction and use of LISA
  sensitivity curves}},\ }\href {https://doi.org/10.1088/1361-6382/ab1101}
  {\bibfield  {journal} {\bibinfo  {journal} {Class. Quant. Grav.}\ }\textbf
  {\bibinfo {volume} {36}},\ \bibinfo {pages} {105011} (\bibinfo {year}
  {2019})},\ \Eprint {https://arxiv.org/abs/1803.01944} {arXiv:1803.01944
  [astro-ph.HE]} \BibitemShut {NoStop}%
\bibitem [{\citenamefont {Janssen}\ \emph {et~al.}(2015)\citenamefont {Janssen}
  \emph {et~al.}}]{Janssen:2014dka}%
  \BibitemOpen
  \bibfield  {author} {\bibinfo {author} {\bibfnamefont {G.}~\bibnamefont
  {Janssen}} \emph {et~al.},\ }\bibfield  {title} {\bibinfo {title}
  {{Gravitational wave astronomy with the SKA}},\ }\href
  {https://doi.org/10.22323/1.215.0037} {\bibfield  {journal} {\bibinfo
  {journal} {PoS}\ }\textbf {\bibinfo {volume} {AASKA14}},\ \bibinfo {pages}
  {037} (\bibinfo {year} {2015})},\ \Eprint {https://arxiv.org/abs/1501.00127}
  {arXiv:1501.00127 [astro-ph.IM]} \BibitemShut {NoStop}%
\bibitem [{\citenamefont {Breitbach}\ \emph {et~al.}(2019)\citenamefont
  {Breitbach}, \citenamefont {Kopp}, \citenamefont {Madge}, \citenamefont
  {Opferkuch},\ and\ \citenamefont {Schwaller}}]{Breitbach:2018ddu}%
  \BibitemOpen
  \bibfield  {author} {\bibinfo {author} {\bibfnamefont {M.}~\bibnamefont
  {Breitbach}}, \bibinfo {author} {\bibfnamefont {J.}~\bibnamefont {Kopp}},
  \bibinfo {author} {\bibfnamefont {E.}~\bibnamefont {Madge}}, \bibinfo
  {author} {\bibfnamefont {T.}~\bibnamefont {Opferkuch}},\ and\ \bibinfo
  {author} {\bibfnamefont {P.}~\bibnamefont {Schwaller}},\ }\bibfield  {title}
  {\bibinfo {title} {{Dark, cold, and noisy: Constraining secluded hidden
  sectors with gravitational waves}},\ }\href
  {https://doi.org/10.1088/1475-7516/2019/07/007} {\bibfield  {journal}
  {\bibinfo  {journal} {J. Cosmol. Astropart. Phys.}\ }\textbf {\bibinfo
  {volume} {2019}}\bibfield  {number} {\bibinfo  {number} { (07)},\ \bibinfo
  {pages} {007}},\ }\Eprint {https://arxiv.org/abs/1811.11175}
  {arXiv:1811.11175 [hep-ph]} \BibitemShut {NoStop}%
\bibitem [{\citenamefont {Sesana}\ \emph {et~al.}(2019)\citenamefont {Sesana}
  \emph {et~al.}}]{Sesana:2019vho}%
  \BibitemOpen
  \bibfield  {author} {\bibinfo {author} {\bibfnamefont {A.}~\bibnamefont
  {Sesana}} \emph {et~al.},\ }\bibfield  {title} {\bibinfo {title} {{Unveiling
  the Gravitational Universe at \textmu Hz Frequencies}},\ }\href@noop {} {\
  (\bibinfo {year} {2019})},\ \Eprint {https://arxiv.org/abs/1908.11391}
  {arXiv:1908.11391 [astro-ph.IM]} \BibitemShut {NoStop}%
\bibitem [{\citenamefont {Arzoumanian}\ \emph {et~al.}(2018)\citenamefont
  {Arzoumanian} \emph {et~al.}}]{Arzoumanian:2018saf}%
  \BibitemOpen
  \bibfield  {author} {\bibinfo {author} {\bibfnamefont {Z.}~\bibnamefont
  {Arzoumanian}} \emph {et~al.} (\bibinfo {collaboration} {NANOGRAV}),\
  }\bibfield  {title} {\bibinfo {title} {{The NANOGrav 11 year data set:
  Pulsar-timing constraints on the stochastic gravitational-wave background}},\
  }\href {https://doi.org/10.3847/1538-4357/aabd3b} {\bibfield  {journal}
  {\bibinfo  {journal} {Astrophys. J.}\ }\textbf {\bibinfo {volume} {859}},\
  \bibinfo {pages} {47} (\bibinfo {year} {2018})},\ \Eprint
  {https://arxiv.org/abs/1801.02617} {arXiv:1801.02617 [astro-ph.HE]}
  \BibitemShut {NoStop}%
\bibitem [{\citenamefont {Arzoumanian}\ \emph {et~al.}(2020)\citenamefont
  {Arzoumanian} \emph {et~al.}}]{Arzoumanian:2020vkk}%
  \BibitemOpen
  \bibfield  {author} {\bibinfo {author} {\bibfnamefont {Z.}~\bibnamefont
  {Arzoumanian}} \emph {et~al.} (\bibinfo {collaboration} {NANOGrav}),\
  }\bibfield  {title} {\bibinfo {title} {{The NANOGrav 12.5 yr data set: Search
  for an isotropic stochastic gravitational-wave background}},\ }\href
  {https://doi.org/10.3847/2041-8213/abd401} {\bibfield  {journal} {\bibinfo
  {journal} {Astrophys. J. Lett.}\ }\textbf {\bibinfo {volume} {905}},\
  \bibinfo {pages} {L34} (\bibinfo {year} {2020})},\ \Eprint
  {https://arxiv.org/abs/2009.04496} {arXiv:2009.04496 [astro-ph.HE]}
  \BibitemShut {NoStop}%
\bibitem [{\citenamefont {Ratzinger}\ and\ \citenamefont
  {Schwaller}(2021)}]{Ratzinger:2020koh}%
  \BibitemOpen
  \bibfield  {author} {\bibinfo {author} {\bibfnamefont {W.}~\bibnamefont
  {Ratzinger}}\ and\ \bibinfo {author} {\bibfnamefont {P.}~\bibnamefont
  {Schwaller}},\ }\bibfield  {title} {\bibinfo {title} {{Whispers from the dark
  side: Confronting light new physics with NANOGrav data}},\ }\href
  {https://doi.org/10.21468/SciPostPhys.10.2.047} {\bibfield  {journal}
  {\bibinfo  {journal} {SciPost Phys.}\ }\textbf {\bibinfo {volume} {10}},\
  \bibinfo {pages} {047} (\bibinfo {year} {2021})},\ \Eprint
  {https://arxiv.org/abs/2009.11875} {arXiv:2009.11875 [astro-ph.CO]}
  \BibitemShut {NoStop}%
\bibitem [{\citenamefont {Caprini}\ \emph {et~al.}(2009)\citenamefont
  {Caprini}, \citenamefont {Durrer}, \citenamefont {Konstandin},\ and\
  \citenamefont {Servant}}]{Caprini:2009fx}%
  \BibitemOpen
  \bibfield  {author} {\bibinfo {author} {\bibfnamefont {C.}~\bibnamefont
  {Caprini}}, \bibinfo {author} {\bibfnamefont {R.}~\bibnamefont {Durrer}},
  \bibinfo {author} {\bibfnamefont {T.}~\bibnamefont {Konstandin}},\ and\
  \bibinfo {author} {\bibfnamefont {G.}~\bibnamefont {Servant}},\ }\bibfield
  {title} {\bibinfo {title} {{General properties of the gravitational wave
  spectrum from phase transitions}},\ }\href
  {https://doi.org/10.1103/PhysRevD.79.083519} {\bibfield  {journal} {\bibinfo
  {journal} {Phys. Rev. D}\ }\textbf {\bibinfo {volume} {79}},\ \bibinfo
  {pages} {083519} (\bibinfo {year} {2009})},\ \Eprint
  {https://arxiv.org/abs/0901.1661} {arXiv:0901.1661 [astro-ph.CO]}
  \BibitemShut {NoStop}%
\bibitem [{\citenamefont {Co}\ \emph {et~al.}(2021)\citenamefont {Co},
  \citenamefont {Harigaya},\ and\ \citenamefont {Pierce}}]{Co:2021rhi}%
  \BibitemOpen
  \bibfield  {author} {\bibinfo {author} {\bibfnamefont {R.~T.}\ \bibnamefont
  {Co}}, \bibinfo {author} {\bibfnamefont {K.}~\bibnamefont {Harigaya}},\ and\
  \bibinfo {author} {\bibfnamefont {A.}~\bibnamefont {Pierce}},\ }\bibfield
  {title} {\bibinfo {title} {{Gravitational waves and dark photon dark matter
  from axion rotations}},\ }\href@noop {} {\  (\bibinfo {year} {2021})},\
  \Eprint {https://arxiv.org/abs/2104.02077} {arXiv:2104.02077 [hep-ph]}
  \BibitemShut {NoStop}%
\end{thebibliography}%

\end{document}